\documentclass[iop,letter]{emulateapj}

\newcommand{\kms}{\,km~s$^{-1}$}    \newcommand{\msun}{\,M$_\odot$}
\newcommand{\os}{\ion{O}{6}}       \newcommand{\sit}{\ion{Si}{3}}     
\newcommand{\hi}{\ion{H}{1}}       \newcommand{\ct}{\ion{C}{3}}
\newcommand{\lya}{Lyman-$\alpha$}  \newcommand{\hw}{\ion{H}{2}}
\newcommand{\cw}{\ion{C}{2}}       \newcommand{\siw}{\ion{Si}{2}}
\newcommand{\mgw}{\ion{Mg}{2}}
\newcommand{\sqcm}{\,cm$^{-2}$}     \newcommand{\hst}{\emph{HST}}
\newcommand{\nos}{23}  
\newcommand{\ndet}{17} 
\newcommand{\ndetword}{Seventeen}

\begin{document}
\shorttitle{LLS high ions and kinematics}
\shortauthors{Fox et al.}
\title{The high-ion content and kinematics of low-redshift Lyman Limit Systems}

\footnotetext[1]{Based on observations taken under programs 
  11508, 11520, 11541, 11598, 11692, 11741, 12025, 12038, and 12466 
  of the NASA/ESA Hubble Space Telescope, obtained at the Space 
  Telescope Science Institute, which is operated by the Association of 
  Universities for Research in Astronomy, Inc., under NASA contract 
  NAS 5-26555.} 
\author{Andrew J. Fox$^1$,  
  Nicolas Lehner$^2$, Jason Tumlinson$^1$, J. Christopher Howk$^2$, 
  Todd M. Tripp$^3$, J. Xavier Prochaska$^4$, John M. O'Meara$^5$, 
  Jessica K. Werk$^4$, Rongmon Bordoloi$^1$, Neal Katz$^3$,
  Benjamin D. Oppenheimer$^{6,7}$, \& Romeel Dav\'e$^8$}
\affil{$^1$ Space Telescope Science Institute, 3700 San Martin Drive,
  Baltimore, MD 21218\\
$^2$ Department of Physics, University of Notre Dame, 225 Nieuwland 
Science Hall, Notre Dame, IN 46556\\
$^3$ Department of Astronomy, University of Massachusetts,
Amherst, MA 01003\\
$^4$ UCO/Lick Observatory, University of California, Santa
Cruz, CA 95064\\
$^5$ Department of Physics, Saint Michael's College, 
One Winooski Park, Colchester, VT 05439\\
$^6$ Leiden Observatory, Leiden University, NL-2300 RA Leiden, Netherlands\\
$^7$ CASA, Department of Astrophysical and Planetary Sciences, University 
of Colorado, Boulder, CO 80309\\
$^8$ University of the Western Cape, Robert Sobukwe Road, Bellville, 7535,
South Africa}

\email{afox@stsci.edu}

\begin{abstract}
We study the high-ion content and kinematics
of the circumgalactic medium (CGM) around low-redshift galaxies 
using a sample of \nos\ Lyman Limit Systems (LLSs) at 
$0.08\!<\!z\!<\!0.93$ observed with the Cosmic Origins Spectrograph 
onboard the \emph{Hubble Space Telescope}. 
In Lehner et al. (2013), we recently showed that low-$z$ LLSs have a 
bimodal metallicity distribution. Here we extend that analysis to
search for differences between the high-ion and kinematic properties
of the metal-poor and metal-rich branches.
We find that metal-rich LLSs tend to show higher \os\ columns 
and broader \os\ profiles than metal-poor LLSs.
The total \hi\ line width ($\Delta v_{90}$ statistic) in LLSs is not 
correlated with metallicity, 
indicating that the \hi\ kinematics alone cannot 
be used to distinguish inflow from outflow and gas recycling.
Among the \ndet\ LLSs with \os\ detections, all but two show evidence
of kinematic sub-structure, in the form of \os--\hi\ centroid offsets,
multiple components, or both.
Using various scenarios for how the metallicity in the high-ion and 
low-ion phases of each LLS compare, we constrain the ionized hydrogen 
column in the \os\ phase to lie in the range log\,$N$(\hw)$\sim$17.6--20.
The \os\ phase of LLSs is a substantial baryon reservoir, with 
$M$(high-ion)$\sim$10$^{8.5-10.9}$($r$/150\,kpc)$^2$\msun, similar to
the mass in the low-ion phase.
Accounting for the \os\ phase approximately doubles the contribution of 
low-$z$ LLSs to the cosmic baryon budget.
\end{abstract}

\section{Introduction}
The extended ionized halos of galaxies play an important role in the 
mechanisms that drive galaxy evolution. Loosely referred to as the 
circumgalactic medium (CGM), these halos serve as reservoirs of 
baryons and metals, channels through which inflowing and outflowing 
gas pass, and shields which can hinder the passage of cool gas
and quench star formation in the host galaxy.
The exchange of matter between galaxies and their CGM is the engine of
galaxy growth \citep{Da12}.
Observationally, the denser regions of the CGM can be 
seen in the form of Lyman Limit Systems (LLSs), optically-thick quasar 
absorbers with \hi\ columns between 10$^{16}$ and 10$^{19}$\sqcm\ 
\citep{Ty82,St90,Ri11a,OM13,Fu13}\footnote{LLSs are often defined
as extending down to 10$^{17}$ or 10$^{17.2}$\sqcm, but in this paper we
include ``partial'' LLSs down to 
$N$(\hi)=10$^{16.0}$\sqcm, because these absorbers still
create a noticeable spectral break at the Lyman Limit.} 
bridging the gap in \hi\ column between
the \lya\ forest at $N$(\hi)$\ll$10$^{16}$\sqcm, and sub-damped \lya\ 
(sub-DLA a.k.a. super-LLS) absorbers at $10^{19}\!<\!N$(\hi)$<$10$^{20.3}$\sqcm\ 
and bona-fide DLAs at $N$(\hi)$>$10$^{20.3}$\sqcm.

The connection between LLSs and the CGM is based on galaxy detections in 
the fields of quasars whose spectra contain LLSs
\citep{Be86, Be94,St95, CP00,Je05,Pr06,Pr11,Le09,St10,St13,Ri11b,Tu11b,Tr11}.
Observations have established that the CGM can also be probed by 
strong \lya\ forest systems down to $\sim$10$^{14.5}$\sqcm\ 
\citep{Pr11, Ru12, Ke13}, so LLSs are not the 
only category of QSO absorber that traces the CGM, but 
nonetheless they offer a convenient means to identify CGM material.
Accurate \hi\ column densities can be obtained in LLSs by
measuring the depth of the Lyman break and the strength of the 
Lyman series absorption lines.
Furthermore, the sensitivity of current ultraviolet (UV) data is such
one needs to study LLSs (as opposed to lower column density systems)
to be sensitive to \emph{low-metallicity} CGM gas.
Simulations reproduce the CGM-LLS connection, predicting that absorbers with 
$N$(\hi)$\ga\!10^{15}$\sqcm\ trace the CGM of galaxies within impact parameters 
of $\approx$300\,kpc \citep[e.g.][]{SH03,Op10,Sm11,Fu11,Si12,Vo12,BF13}.

In \citet[][hereafter L13]{Le13}, we used a sample of 28 
low-redshift ($z\!<\!1$) LLSs observed with the Cosmic Origins Spectrograph
(COS) onboard \hst\ to show that LLSs have a bimodal metallicity distribution,
with one peak near 2.5\% solar and the other near 50\% solar metallicity,
and a similar number of absorbers in each branch.
A natural interpretation of this result is that the low-metallicity
systems trace primitive gas accreting onto galaxies 
whereas the high-metallicity systems trace enriched outflowing or 
recycling gas (L13). If this interpretation is correct, then
the covering fraction of inflowing and outflowing/recycling gas 
in the LLS \hi\ column density range 
is approximately equal. On the surface, this seems surprising
given that the infall of cold gas onto galaxies is found in simulations 
to be filamentary, generally covering a small solid angle 
\citep{Oc08, Ke09, FG11, Fu11, Mu12, VV12, Jo12b, Sh13}. However, the inflow
covering fraction depends on halo mass \citep{St11a}, and \hi\ 
columns may be lower in outflows where the gas is hot and/or highly ionized.
Therefore, the full implications of the bimodality of LLS metallicities on 
the prevalence of inflow and outflow is unclear.

The LLS metallicities reported in L13 were derived from measurements of 
\hi\ and a range of low- and intermediate-ionization UV metal lines 
(singly- and doubly-ionized species), together with 
photoionization modeling to calculate the ionization corrections. 
However, these photoionization models generally produce only negligible 
amounts of \os, despite the detection of this line in absorption
in many LLSs at low redshift 
\citep[$z\!<\!1$;][]{CP00,Co08,Tr08,Le09,CM09, Sa10,Tu11b,Ka12,Cr13a},
intermediate redshift \citep[$z\!\approx\!1$;][]{Di03,Zo04},
and high redshift \citep[$z\!>\!2$;][Lehner et al. 2014, 
in prep.]{KT97,KT99,Si02}. 
\os\ absorption is also seen in some sub-DLAs \citep{Fo07,Cr13a,Cr13b}, 
which trace the CGM 
just like their lower-$N$(\hi) cousins \citep{Tr05, Ba12}.
Until now, the relationship between the low-ion and high-ion phases of LLSs 
has not been explored in a large sample at any redshift.

In this paper we focus on the high-ionization phase of low-$z$ LLSs by 
surveying absorption in the \os\ 1031, 1037 doublet in the L13 sample. 
We also measure and discuss their kinematic properties.
The overall goal is to explore whether the bimodality of LLSs
seen in their metallicities is evident in any of their other properties.

Since 113.9\,eV is required to ionize 
\ion{O}{5} to \os\ (O$^{+4}$ to O$^{+5}$), \os\ is a tracer 
of either hot, collisionally ionized gas at a few times 10$^5$\,K
or warm photoionized plasma at $\sim$10$^4$\,K subject 
to a hard (non-stellar) ionizing radiation field.
Its properties have been studied in a wide range of interstellar, 
circumgalactic, and intergalactic environments 
\citep[see summaries in][]{He02, Fo11} and 
are explored in detail in numerical simulations 
\citep{OD09, CC11, TG11, Sm11}.
\os\ has recently been surveyed in the CGM of 42 low-$z$ galaxies
in the COS-Halos sample \citep{Tu11a,Tu13}, with the key result
that the \os\ column is strongly dependent on the presence
or absence of star formation of the host galaxy; 
our approach here is complementary 
in that we are using an \hi\ selection rather than a galaxy selection, 
but we are almost certainly probing the CGM in each case. 

This paper is structured as follows. In \S2 we briefly describe the sample 
selection. In \S3 we discuss our measurement procedures. In \S4 we present 
results on \os\ column density, LLS kinematics, and \hw\ column in 
the \os\ phase. We present a discussion in \S5 and a summary in \S6.
Throughout this paper we use atomic data for absorption lines from 
\citet{Mo03} and \citet{Ve94} and the solar reference oxygen abundance 
log\,(O/H)$_\odot$=--3.31 from \citet{As09}.

\section{Sample Selection}
We began with the L13 sample of 28 LLSs with \hi\ columns
between 10$^{16.11}$ and 10$^{17.00}$\sqcm\ and redshifts between
0.081 and 0.927, all selected by \hi\ absorption and observed by \hst.
The \hi\ selection (as opposed to a \mgw\ selection or 
other metal-line selection) is important for ensuring that no bias 
toward metal-rich systems is introduced.
These systems are drawn from a range of \hst\ programs with a variety of 
science goals, and in all cases the quasars were not targeted to show LLSs.
Some sightlines were chosen to 
\emph{avoid} optically thick absorbers with log\,$N$(\hi)$\gg$17.
Seven of the systems come from the COS-Halos dataset, but these seven
do not overlap with the absorption systems with associated galaxies 
\citep{Tu11a, We13}.
Measurements of the properties of the low-ion 
phase of these 28 systems are given in Table 4 of L13. 
We also make use of supplemental high-resolution ($\approx$6\kms\ FWHM) 
Keck/HIRES observations covering \mgw\ $\lambda\lambda$ 2796, 2803 
absorption in eleven of the LLSs, and medium-resolution ($\approx$55\kms\ FWHM) 
Magellan/MagE observations of \mgw\ in two of the LLSs. These data were
reduced as described in \citet{Th11} and \citet{We12}, respectively.

We then selected those systems with COS data covering the \os\ 
1031, 1037 doublet, giving \nos\ LLSs, which is the primary sample 
for this paper (see Table 1).
Twelve of these are in the metal-poor branch ([Z/H]$\le$--1)
and eleven are in the metal-rich branch ([Z/H]$>$--1).
All the spectra have a velocity resolution of $\approx$18\kms\ (FWHM),
and were taken with either the G130M, G160M, or (in one case) G185M gratings. 
The data were reduced and normalized following the procedures described in 
\citet{Th11}, \citet{Me11}, and L13.
This includes determining and correcting for 
wavelength shifts between individual
exposures using Galactic interstellar absorption lines.
The data were rebinned by three native pixels, to give final spectra with 
two pixels per 18\kms\ resolution element.
Details of the COS instrument are given in \citet{Gr12}.

Velocity stacks showing the absorption profiles of \hi, a low ion 
(\cw\ or \siw), an intermediate ion (\ct\ or \sit), and the high-ion 
\os\ are presented for each LLS in the sample in the Appendix.
Either \cw\ $\lambda$1036, \cw\ $\lambda$903.96, or \cw\ $\lambda$1334 is
chosen to represent the low ions, since these are three of the strongest 
available lines, although we use \siw\ $\lambda$1020 or \siw\ $\lambda$1260 
instead in cases where the \cw\ lines are blended or not covered.
\cw\ $\lambda$1036 is adopted as the default low-ion line
since it lies in-between the two members of the \os\ $\lambda\lambda$ 
1031, 1037 doublet, and hence is unlikely to suffer from zero-point 
offsets.
Likewise, we use \ct\ $\lambda$977 as the default intermediate ion,
but replace it with \sit\ $\lambda$1206 when \ct\ is saturated or not covered.
For \hi, we adopt an unsaturated (usually high-order) Lyman series line 
whenever possible. These stacks allow for comparison of
the absorption in each phase of the system, and show the data on which
our measurements were made. Full velocity-profile plots showing all available
metal lines in many (but not all) of these systems are given in L13. 

\section{Measurements} 
For each LLS, we measure the absorption in each line of interest using
the apparent optical depth (AOD) technique of \citet{SS91}.
In this technique, the AOD in each pixel is defined as
$\tau_a(v)$=ln\,$[F_c(v)/F(v)]$,
where $F_c(v)$ is the continuum level and $F(v)$ is the 
observed flux. The apparent column density in each pixel is 
$N_a(v)= (m_e c/\pi e^2)[\tau_a(v)/f\lambda] = 3.768
\times 10^{14}\,[\tau_a(v)/f\lambda]$ \sqcm\,(km\,s$^{-1}$)$^{-1}$,
where $f$ is the oscillator strength of the transition and $\lambda$ is
the wavelength in Angstroms. Integrating over the profile
gives the apparent column density, $N_a=\int_{v_{\rm min}}^{v_{\rm max}}N_a(v){\rm d}v$,
where $v_{\rm min}$ and $v_{\rm max}$ define the velocity range of absorption.

Apparent column density profiles for each system in the sample are 
given in the Appendix. In each LLS we fit Gaussian components to the 
strongest absorption feature in the \hi, \cw, \ct\ or \sit, and 
\os\ profiles, and measure the velocity centroid offsets 
$\delta v_0$(\os--\hi), $\delta v_0$(\ct--\hi), and
$\delta v_0$(\cw--\hi).
We note that a large centroid offset $\delta v_0$(\os--\hi)
does not necessarily indicate the \hi\
and \os\ phases are unrelated; it can also indicate a multi-component 
absorbers with \hi/\os\ ratios that vary from component to component. 
Nonetheless, the statistic illustrates whether the bulk of the \hi\ 
and the bulk of the \os\ are offset in velocity, which is a useful indicator 
of the distribution of the two ions in the absorber. 

To verify whether a given offset is significant, one needs to know the 
reliability of the wavelength solution. The nominal accuracy of the COS 
wavelength solution is $\approx$15\kms\ 
\citep[COS Instrument Handbook;][]{Ho12}.
However, we can verify the accuracy on a case-by-case basis by checking 
the alignment of multiple low-ionization lines from the same system; if the 
centroids of all the low ions agree to within a few\kms\ (which they 
usually do), but \os\ is offset by several times that amount, we can be 
confident the offset is real, especially since our default low-ion line is
\cw\ $\lambda$1036, which lies in-between the two lines of the \os\ 
$\lambda\lambda$1031,1037 doublet, and thus offers an excellent reference
because it falls in the same regions of the COS detector.

In addition to the velocity centroid offset, we also analyze the total 
velocity width of absorption $\Delta v_{90}$, which we measure for 
\os, \cw, \mgw, \ct\ or \sit, and \hi.
This width is defined as the velocity interval containing the central 
90\% of the integrated apparent optical depth in the line \citep{PW97}.
It is measured by integrating each line between $v_{\rm min}$ and $v_{\rm max}$
and finding the difference in velocity between the pixels at which 5\% and 95\%
of the total optical depth has been reached. For each LLS, $v_{\rm min}$ and 
$v_{\rm max}$ are selected by eye based on assessing the absorption in all ions 
in the system, but are adjusted in cases of contamination. $\Delta v_{90}$ has 
the advantage of being unaffected by the presence of weak satellite 
components, which do not contribute significantly to the total optical depth 
in the line, and is thus preferable to the alternative measure of total
line width $v_{\rm max}$--$v_{\rm min}$. However, $\Delta v_{90}$ does have
limitations:
for a saturated line, only an upper limit on $\Delta v_{90}$ can be derived, 
whereas for a weak line, part of the absorption can be missed in the noise.
To deal with these limitations, one approach \citep[see][]{Ha98, Le06} 
is to restrict $\Delta v_{90}$ measurements to lines where
$0.1\!<\!F(v)/F_c(v)\!<\!0.6$, and we follow this where possible when choosing
between \ct\ and \sit, and when selecting which \hi\ Lyman series line to 
measure.

The measurements analyzed in this paper are given in Tables 1 and 2. 

\section{Results}

\subsection{\os\ Column Densities}
\os\ absorption is detected at 3$\sigma$ significance or above
in \ndet\ of the \nos\ LLSs in the sample; 
among these detections, the mean and standard deviation of log\,$N$(\os) is 
14.22$\pm$0.44. 
For comparison, this is lower than the mean \os\ column ($\approx$14.5) 
observed in sight lines through the halos of star-forming galaxies
at similar redshifts at impact parameters less than 150\,kpc \citep{Tu11a},
and less than the average \os\ column seen in sightlines
though the Milky Way halo integrated over velocity
\citep[$\approx$14.4;][]{Wa03, Sa03}.
In the six LLSs with \os\ non-detections, the 3$\sigma$ upper
limits on the \os\ column range from log\,$N$(\os)$<$13.29 to $<$14.14; 
two of these non-detections lie within the distribution 
of $N$(\os) from the \os\ detections.

\begin{figure}[!ht]
\epsscale{1.2}\plotone{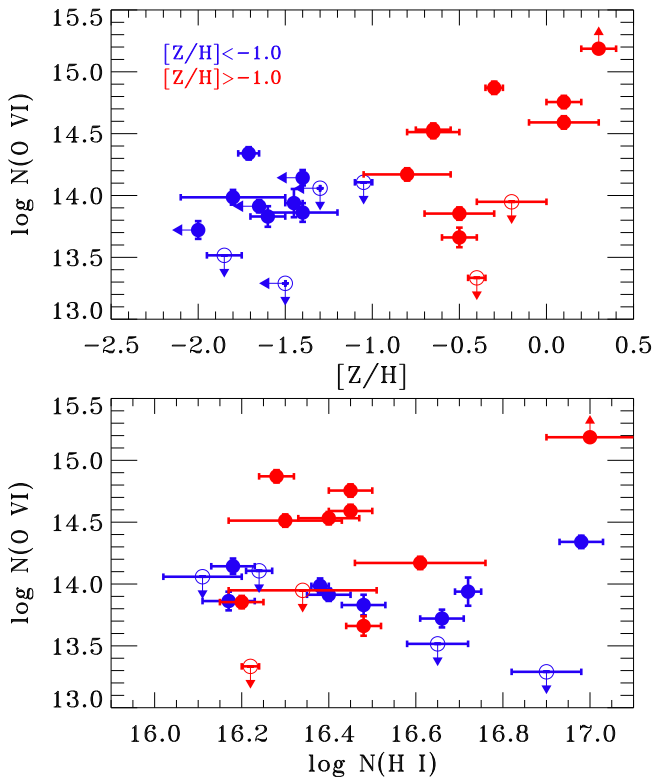}
\caption{{\bf Top panel:} \os\ column as a function of metallicity for the 
\nos\ LLSs in our sample. The metallicities are taken from L13 and are derived 
from applying photoionization models to the observed low-ion column densities.
\os\ detections are shown with filled circles and non-detections with open 
circles and attached upper limit arrows.
{\bf Lower panel:} \os\ column vs \hi\ column. All our systems have 
16$<$log\,$N$(\hi)$<$17 and so fall at the lower end of the LLS column 
density range. Blue points show low-metallicity LLSs ([Z/H]$\le$--1)
and red points show high-metallicity LLSs ([Z/H]$>$--1).}
\end{figure}

\begin{figure}[!ht]
\epsscale{1.2}\plotone{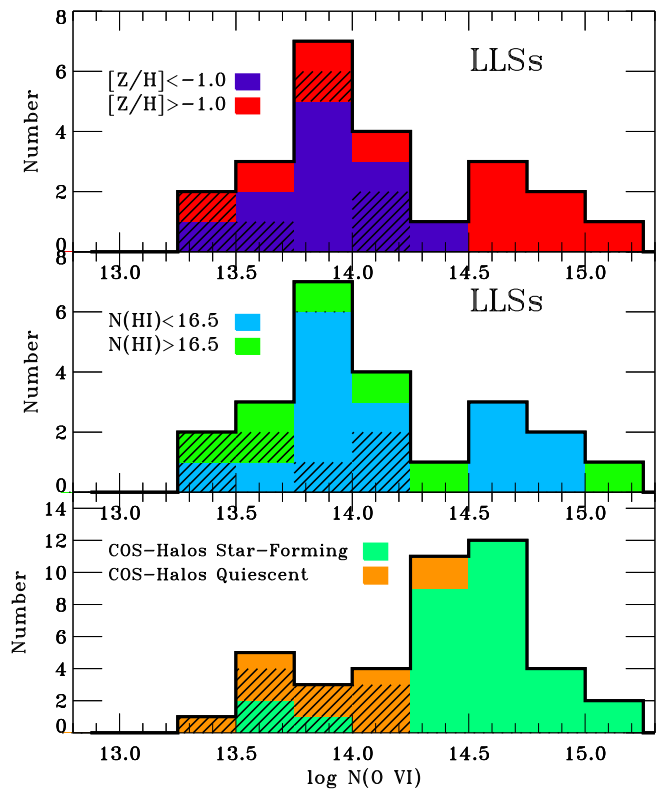}
\caption{{\bf Top panel:} histograms of \os\ column density 
in the full LLS sample (black line),
low-metallicity sample ([X/H]$\le$--1.0; blue), and 
high-metallicity sample ([X/H]$>$--1.0; red).
Upper limits are shown with hatched regions.
{\bf Middle panel:} same as the top panel except the sample is divided 
into low-\hi\ column (log\,$N$(\hi)$<$16.5; blue) and high-\hi\ column 
(log\,$N$(\hi)$>$16.5; green) sub-samples.
{\bf Bottom panel:} distributions of $N$(\os) in the CGM of
the COS-Halos sample of star-forming (green) and quiescent (orange) galaxies 
\citep{Tu11a}. Note that many of these COS-Halos absorbers 
have log\,$N$(\hi)$<$16 and hence are not LLSs.}
\end{figure}

In the top panel of Figure 1 we plot the \os\ column density against 
metallicity, where the metallicity is derived in the low-ion phase of the 
LLS by L13. A clear trend is seen in which the high-metallicity LLSs  
tend to show stronger \os\ than the low-metallicity LLSs. 
All of the 12 metal-poor LLSs ([Z/H]$\le$--1) show weak \os\
[log\,$N$(\os)$<$14.4], 
whereas 6 of the 11 metal-rich LLSs ([Z/H]$>$--1) show 
strong \os\ [log\,$N$(\os)$>$14.4]. 
A Kendall-tau correlation test finds evidence for a weak correlation 
between $N$(\os) and [Z/H] (correlation coefficient=0.35) significant at 
the 97.9\% level when treating the upper limits as data points. If we ignore
the upper limits and redo the Kendall tau analysis, the significance of 
the correlation remains similar at 97.6\%.

The $N$(\os)-metallicity trend can also be seen in the top panel of 
Figure 2, where we show the distributions of $N$(\os) for the 
[Z/H]$\le$--1 and [Z/H]$>$--1 sub-samples. A two-sided 
Kolmogorov-Smirnov (K-S) test rules out the hypothesis that the 
two distributions are drawn from the same parent population at the 
96.4\% confidence level 
(this rises to 97.6\% confidence when ignoring the \os\ non-detections). 
In the bottom panel of Figure 2 we display the COS-Halos $N$(\os) distribution,
showing the \os\ columns measured in the CGM of star-forming galaxies (SFGs)
and quiescent galaxies at impact parameters up to 150\,kpc
\citep{Tu11a,Tu13}, although caution is needed in the comparison since
$\approx$70\% of these systems have 
14.5$<$log\,$N$(\hi)$<$16.0 and hence are not LLSs (L13).
We note that the metal-rich-LLS distribution 
overlaps in the range of $N$(\os) with the COS-Halos
SFG distribution, with a majority of absorbers showing log\,$N$(\os) in the 
range 14--15 in each case, although the COS-Halos SFG distribution is 
much more peaked around its maximum value.

In the lower panel of Figure 1, we directly compare the \os\ and 
\hi\ columns in the LLSs. Although there is a cluster of LLSs with strong 
\os\ and log\,$N$(\hi) lying between 16.2 and 16.5, which creates 
the visual impression of an anti-correlation, no statistically significant 
difference in \os\ column is seen between the low-\hi\ column 
[log\,$N$(\hi)$<$16.5] and high-\hi\ column 
[log\,$N$(\hi)$>$16.5] halves of the samples 
(see also Figure 2, middle panel).

\subsection{LLS Kinematics}
In Figure 3, we compare the velocity width $\Delta v_{90}$ with metallicity,
for five species: \hi, \cw, \mgw, \ct\ or \sit, and \os. 
There is no significant difference in the distributions of $\Delta v_{90}$ 
between the low-metallicity and high-metallicity sub-samples of LLSs 
for either \hi\ or \ct.
For \cw, we cannot reliably compare $\Delta v_{90}$ between the metal-poor and 
metal-rich branches, because there is only one \cw\ detection in the 
metal-poor branch (no limit on $\Delta v_{90}$ can be made when a line 
is not detected.) For \mgw, the metal-rich LLSs generally appear broader
than the metal-poor LLSs, but the sample is small, and there are 
two metal-rich LLSs with narrow \mgw\ ($\Delta v_{90}\!\la\!30$\kms).

The low-ion velocity widths are in general small 
(sub-virial), with mean values $\langle\Delta v_{90}\rangle$
of 98$\pm$12\kms\ for \hi, 65$\pm$8\kms\ for \cw, 
41$\pm$8\kms\ for \mgw, and 83$\pm$14\kms\ for \ct, 
where the uncertainties quoted are the standard errors of the mean,
and where upper limits (from saturated lines) were ignored when 
calculating these averages.
The dispersions around these mean values, $\sigma(\Delta v_{90})$,
are 49\kms\ for \hi, 24\kms\ for \cw, 24\kms\ for \mgw,
and 51\kms\ for \ct. 
Aside from the $z$=0.9270 LLS toward \object{PG1206+459},
for which saturation prevents a measurement of $\Delta v_{90}$ for
most of the lines of interest, the maximum observed $\Delta v_{90}$ is 
215\kms\ for \hi, 238\kms\ for \ct, and 102\kms\ for \cw. 

\begin{figure}
\epsscale{1.2}\plotone{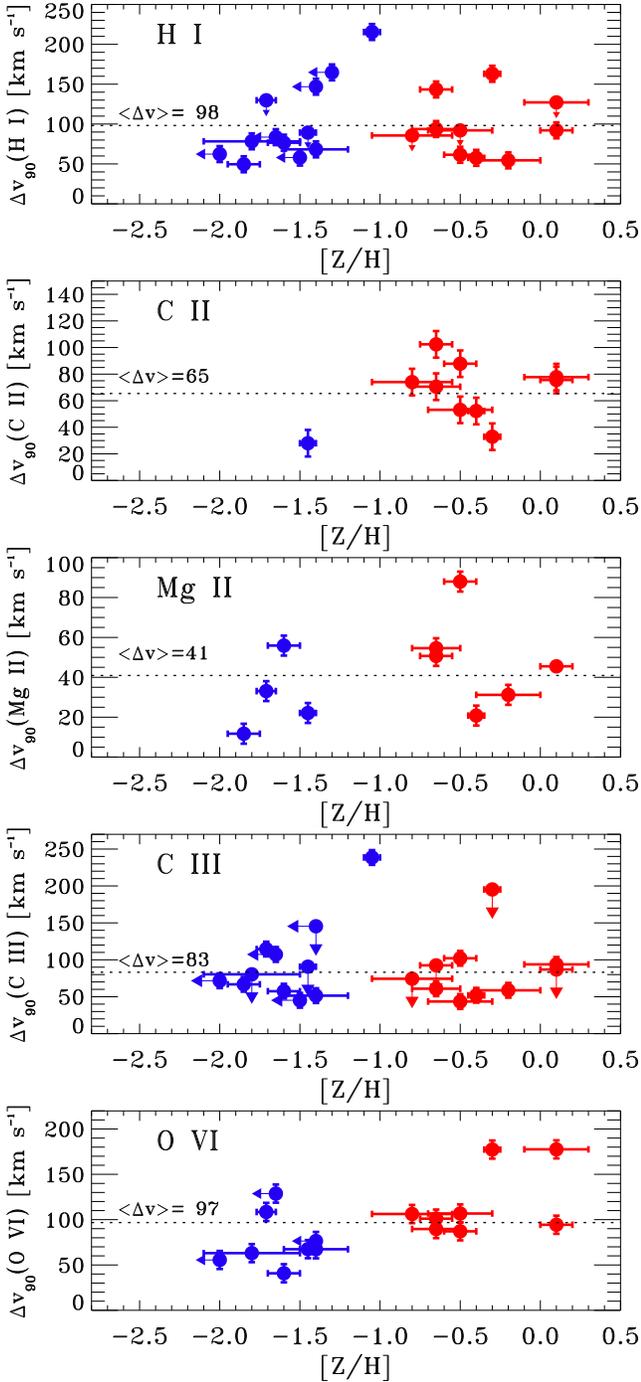}
\caption{Comparison of velocity width $\Delta v_{90}$ with metallicity
for \hi, \cw, \mgw, \ct, and \os. Blue points indicate
low-metallicity LLSs and red points indicate high-metallicity LLSs.
We use \sit\ as a proxy for \ct\ in eight cases where it gives a better 
indication of the intermediate-ion kinematics. 
Saturated lines give upper limits on $\Delta v_{90}$. 
The $z$=0.927 LLS toward PG1206+459 is omitted from this plot;
it shows a large velocity spread, covering $\approx$600\kms.
The sample size varies between ions since $\Delta v_{90}$ can only 
be measured if the line is detected. 
\cw\ is only detected in one of the metal-poor LLSs.
In each panel the mean velocity width is annotated and shown 
with a dotted horizontal line.
Note that each panel has a different scale on the y-axis.} 
\end{figure}

For \os\ (Figure 3, bottom panel),
the mean $\Delta v_{90}$(\os)=97$\pm$10\kms\ is statistically equal to the
mean $\Delta v_{90}$ for \hi, and a factor of $\approx$1.4 higher than 
the mean $\Delta v_{90}$ for \ct, although there is a large 
dispersion in $\Delta v_{90}$ for each line. 
However, the difference with \os\ is that $\Delta v_{90}$ 
is a function of metallicity.
Three of the four systems with the broadest \os\footnote{These are 
the $z$=0.6153 LLS toward \object{HE0439--5254},
the $z$=0.9270 LLS toward \object{PG1206+459}, and the 
$z$=0.1672 absorber toward \object{PKS0405--123}.}
are three of the four most metal-rich LLSs in the sample
(and in these cases, the \os\ absorption is substantially broader than 
the \cw). A Kendall-tau correlation test finds evidence for a 
moderate correlation (correlation coefficient=0.46) between
$\Delta v_{90}$(\os) and [Z/H] significant at the 99.0\% level. 
A two-sided K-S test rules out the hypothesis that the distributions 
of $\Delta v_{90}$(\os) in the metal-poor and metal-rich branches are 
drawn from the same parent population at the 99.3\% confidence level. 

The other kinematic statistic we analyze is the velocity centroid
offset $|\delta v_0|$ between the strongest absorption component 
seen in various ions. 
We find $|\delta v_0|$(\os--\hi) is $>$10\kms\ in 8 of the \ndet\ LLSs
where \os\ is detected, and $>$20\kms\ in 6 of these systems (Figure 4). 
For intermediate ions \ct\ or \sit, denoted \ion{X}{3} for short, we find 
$|\delta v_0|$(\ion{X}{3}--\hi) is $>$10\kms\ in 8 of 
the 20 cases where \ct\ or \sit\ is detected. 
As mentioned in \S3, these centroid offsets provide important 
observational information on the distribution of the absorbing ions in 
velocity space in each absorber, but can reflect component structure 
within the absorber. Large values of $|\delta v_0|$(\os--\hi) indicate 
the bulk of the \os\ is offset from the bulk of the \hi.
The \os--\hi\ centroid offsets distribute similarly for metal-rich and 
metal-poor LLSs (Figure 4), with offsets extending to $\pm$50\kms\
around the systemic redshift of the absorber, and with
no evidence for asymmetry around zero. 

\begin{figure}
\epsscale{1.2}\plotone{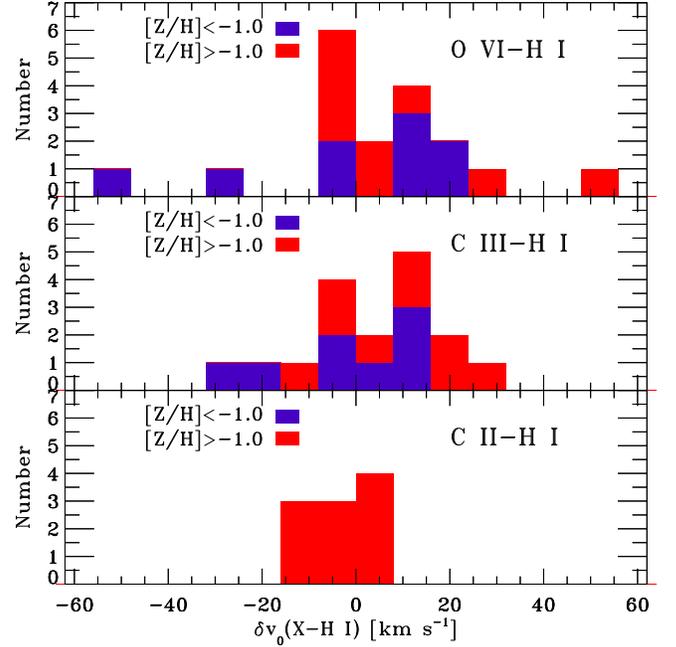}
\caption{Histograms of velocity centroid offset 
in the low-metallicity ([X/H]$<$--1.0; blue) and 
high-metallicity sub-samples ([X/H]$>$--1.0; red), for \os--\hi\ (top),
\ct--\hi\ (middle), and \cw--\hi\ (bottom).
This statistic measures the offset between the strongest
component of absorption in each ion.}
\end{figure}

\begin{figure}
\epsscale{1.2}\plotone{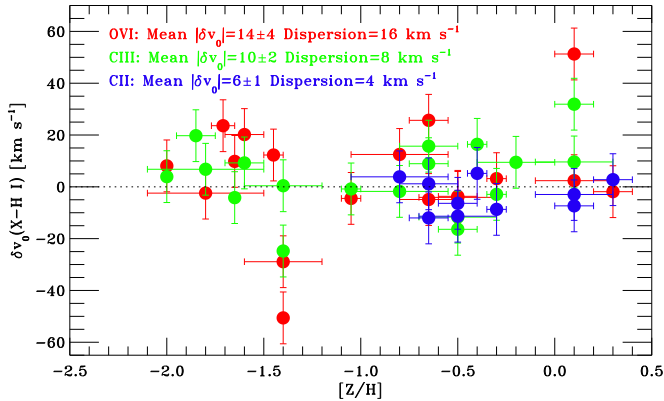}
\caption{Comparison of velocity centroid offset $|\delta v_0|$(X--\hi)
with metallicity where X is \os, \ct, and \cw. 
Both the mean absolute offset and the dispersion in the 
absolute offset increase from \cw\ (blue) to \ct\ (green) to \os\ (red). 
These values are annotated on the plot.}
\end{figure}

The offsets $|\delta v_0|$(X--\hi) where X is
\os, \ct, and \cw\ are plotted against LLS metallicity in Figure 5.
Across the whole sample, we find that the mean absolute offset
$\langle|\delta v_0|\rangle$(X--\hi) is 14$\pm$4\kms\ for \os, 
10$\pm$2\kms\ for \ct, and 6$\pm$1\kms\ for \cw, 
where the uncertainties quoted are the standard errors of the mean. 
The dispersion in the centroid offsets $\sigma(|\delta v_0|)$(X--\hi)
decreases from 15\kms\ for \os--\hi, to 8\kms\ for \ct--\hi, to 
3\kms\ for \cw--\hi.
Therefore, both the mean absolute offset and its dispersion appear to
increase with ionization potential from \cw\ to \ct\ to \os.
This can also be seen in the histograms shown in Figure 4.

It is useful to assess whether the line profiles in each LLS are 
consistent with a single component, i.e. whether the 
\hi\ and all metal lines can co-exist in a single cloud at a single velocity. 
Of the \nos\ LLSs in our sample with \os\ detections, twelve show 
multi-component structure (this structure is most visible in \ct\ 
$\lambda$977, but is often seen in \hi\ and \os\ as well).
The remaining five have absorption profiles that are consistent with a single 
component. Only two absorbers in the sample,
the $z$=0.2694 LLS toward \object{J1619+3342} and the
the $z$=0.7292 LLS toward \object{PG1522+101},
show both a single component \emph{and} a
small centroid offset $|\delta v_0|$(\os--\hi)$<$10\kms\
(these two are both metal-poor LLSs with very weak \os).
In other words, these are the only two systems where the line profiles
allow a solution in which all observed species could arise in a single 
cloud at a single central velocity (and even in these two, the aligned 
\os\ and \hi\ components could arise in a chance coincidence 
of distinct clouds.)

The overall conclusion from our kinematic analysis, considering the total
line widths, the velocity centroid offsets, and the component structure,
is that LLS absorption profiles are complex, with 
differences between the high-ion and low-ion profiles
in all but two of the systems in our sample.
The \os\ profiles show a broader mean $\Delta v_{90}$ than the low-ion 
profiles, and a larger mean centroid offset from \hi\ than the low ions.
In almost half the sample (8 of 17 LLSs) 
show a velocity centroid offset $|\delta v_0|$(\os--\hi)$>$10\kms,
indicating the bulk of the \os\ and \hi\ do not co-exist in the same regions.

\subsection{Ionized Hydrogen Column in \os\ Phase}
The correlation between \os\ column and metallicity reported in \S4.1
has implications for the ionized hydrogen column density $N$(high-ion \hw) 
in the \os\ phase, which is given by:\\

$N$(high-ion \hw)=$N$(\os)/[$f$(\os)(O/H)$_{\rm high-ion}$].\\

Here $f$(\os)$\equiv$\os/O is the \os\ ionization fraction
and (O/H)$_{\rm high-ion}$ is the oxygen abundance in the \os-bearing gas.
To evaluate $N$(high-ion \hw) in each LLS, we need to determine
the ionization level and metallicity.
First, we adopt a common ionization fraction for all systems, using the 
maximum \os\ fraction $f$(\os)$_{\rm max}$=0.2 permitted under 
either equilibrium or non-equilibrium collisional ionization models 
\citep{SD93, GS07, OS13}, which gives a \emph{minimum} high-ion \hw\ column.
To evaluate the metallicity, we consider three scenarios, that cover
the plausible range of parameter space:\\
Case 1: the metallicity in the high-ion phase of each LLS is equal 
to that measured in the low-ion phase. 
This scenario would apply if the low-ion phase condensed out of the 
high-ion phase \citep[see][]{MB04, Jo12a}, or conversely if the high-ion 
phase was heated and evaporated out of the low-ion phase.\\
Case 2: the metallicity in the high-ion phase of each LLS is 0.5\,dex
(factor of three) \emph{higher} in the \os\ phase than the 
low-ion phase, as was found observationally for the LLS at 
$z$=0.2261 toward \object{HE0153--4520}, discussed in detail 
in \citet{Sa11}. By decomposing the \lya\ absorption profile in this system
into narrow (low-ion) and broad (high-ion) components, these authors derived 
the metal abundance separately in the two phases, finding 
[X/H]$_{\rm low-ion}$=--0.8$^{+0.3}_{-0.2}$, 
and [O/H]$_{\rm high-ion}$=--0.28$^{+0.09}_{-0.08}$, 
corresponding to a factor of three enhancement in 
metallicity in the high-ion phase.\\
Case 3: the \os\ phase traces gas with solar metallicity,
as might apply if it traced galactic outflows or recycled wind 
material \citep[e.g.][]{St12}.

The values of $N$(high-ion \hw) evaluated in these three cases are shown 
in Figure 6. The data points show the values calculated in case (1), with the 
downward-pointing ``error'' bars extended to account for cases (2) and (3).
In case 1, log\,$N$(high-ion \hw) lies in the range $\approx$18--20, whereas
in case 3, log\,$N$(high-ion \hw) lies in the narrower range 
$\approx$17.6--18.6.
In cases 1 and 2, $N$(high-ion \hw) is \emph{anti-correlated}
with metallicity, so that the high-metallicity systems show a deficit 
of high-ion plasma even though they tend to show stronger \os.
In other words, in these scenarios
the $N$(\os)--metallicity correlation shown in Figure 1
is shallower than would be expected if driven by metallicity alone.
However, in case 3, $N$(high-ion \hw) shows much less dependence on [Z/H].

\begin{figure}
\epsscale{1.2}\plotone{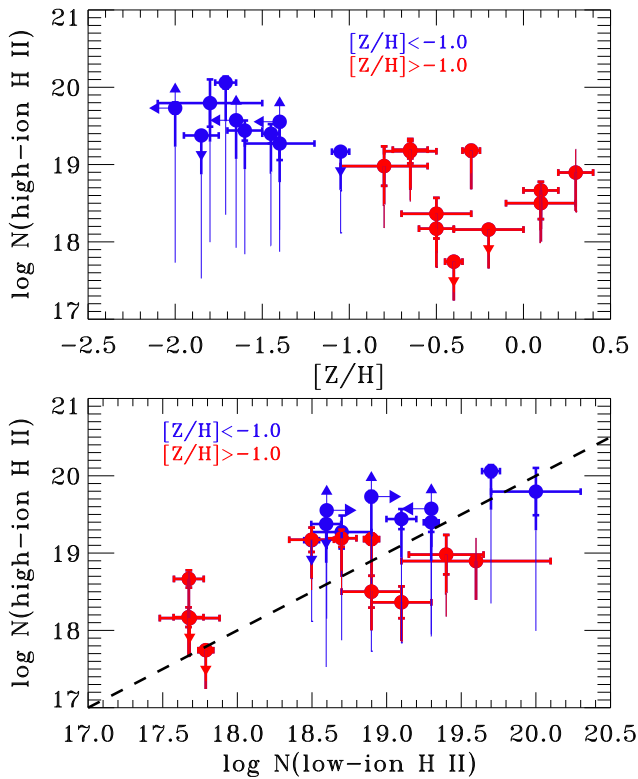}
\caption{{\bf Top panel:} \hw\ column in the \os\ phase 
(=high-ion \hw) as a function of LLS metallicity,
assuming that the metallicity in the low-ion and high-ion phases of 
each LLS are equal (case 1), and that the \os\ ionization 
fraction $f$(\os)=0.2. The thick downward error bars indicate the effect 
of using a 0.5\,dex higher metallicity in the \os\ phase (case 2). 
The thin downward error bars indicate the effect of using a solar 
metallicity in the \os\ phase (case 3).
{\bf Lower panel:} $N$(high-ion \hw) vs $N$(low-ion \hw), 
where the values of $N$(low-ion \hw) are taken from the photoionization 
models of L13.
The dashed line shows the relation $N$(high-ion \hw)=$N$(low-ion \hw). 
Blue and red points show the metal-poor and metal-rich sub-samples.
Lower limits on $N$(high-ion \hw) indicate absorbers with upper limits on 
metallicity. Upper limits on $N$(high-ion \hw) indicate absorbers with no \os\ 
detections.} 
\end{figure}

The high-ion \hw\ columns can be directly compared to the low-ion 
($\approx$10$^4$\,K) \hw\ columns derived in each LLS from the \emph{Cloudy} 
photoionization models of L13. This is shown in the lower panel of Figure 6.
The key result here is that the H$^+$ columns in the high-ion and low-ions 
phases are of similar order, \emph{even when accounting for the possibility 
that (O/H)$_{\rm high-ion}\!\approx\!3$(O/H)$_{\rm low-ion}$} as discussed 
in case 2 above. In addition, the \hw\ columns in the two phases are correlated
(in cases 1 and 2), though this is driven by their similar dependence on 
metallicity, with both being proportional to 1/(Z/H).
Other examples exist in the literature of 
LLSs where $N$(high-ion \hw)$>\!N$(low-ion \hw) \citep{Tr11, Cr13b}.
Intergalactic absorption-line systems with lower \hi\ columns can also show 
this property \citep{Me13}.

The magnitude of the baryon reservoir in the various forms of hydrogen
in LLSs is shown in Figure 7, where the breakdown of 
hydrogen column in each LLS into \hi, low-ion \hw, and high-ion \hw\ is 
plotted for two of our three metallicity cases.
The strength of the green and red 
bars relative to the blue bars clearly indicates that \hi\ is a trace 
constituent of LLSs and \hw\ dominates their baryon content, 
with both low-ion \hw\ and high-ion \hw\ making significant contributions.

\begin{figure*}
\epsscale{1.2}\plotone{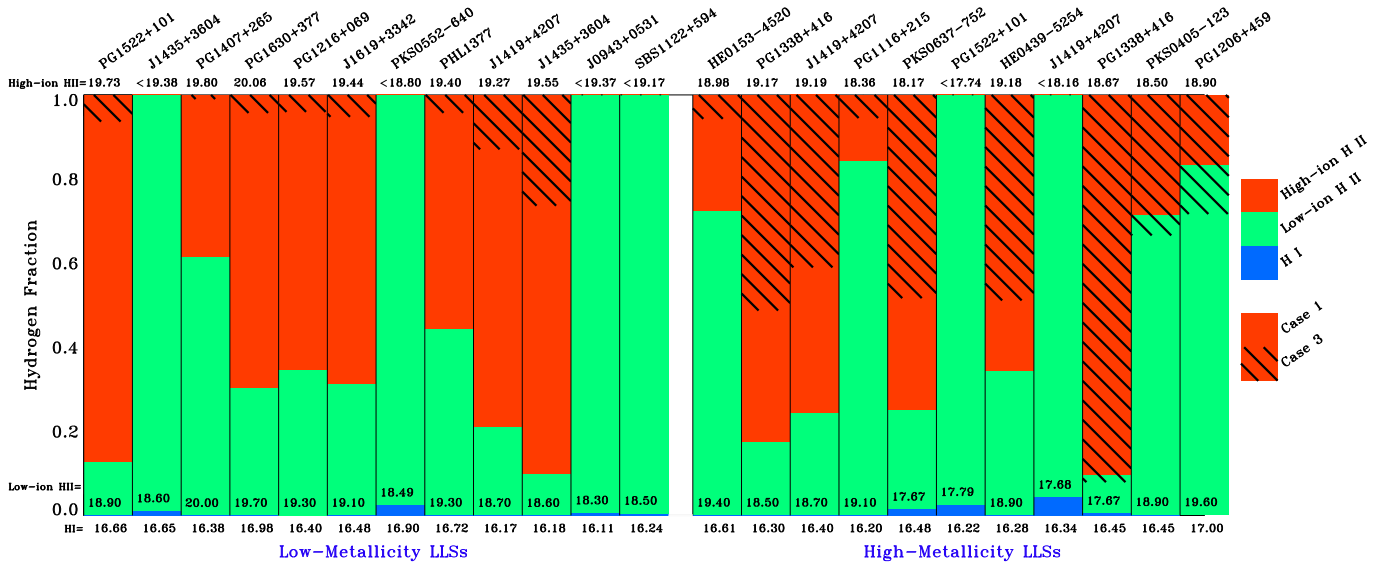}
\caption{Summary of the hydrogen ionization breakdown in each absorber,
illustrating the relative column densities in each phase.
Each column represents a LLS, identified at the top, ordered from
low to high metallicity. The colored bars
represent the fraction of the total hydrogen column residing in the \hi, 
low-ion \hw, and high-ion \hw\ phase. The numbers annotated on each bar show 
the logarithm of the hydrogen column density in that phase. 
The high-ion \hw\ fractions are shown for two of the metallicity cases 
described in \S4.3:
case 1 (full red) where the metallicity in the high-ion and low-ion phase
of each LLS is the same, and 
case 3 (hatched red), where the high-ion metallicity is solar
(case 2 lies in-between). The low-ion \hw\ columns are taken from L13.
In most cases, the \hi\ contribution is negligible since the LLSs are
so highly ionized, and the blue \hi\ bar is not visible.}
\end{figure*}

\section{Discussion}
The recent result of L13 that low-redshift LLSs have a bimodal 
metallicity distribution has placed important constraints 
on the distribution of gas in the CGM of low-redshift galaxies.
In this paper, we have searched for evidence of bimodality in other 
LLS properties, particularly their kinematics and high-ion columns. 
Interestingly, the LLS bimodality is \emph{not} reflected in their 
bulk \hi\ kinematics or intermediate-ion (\ct\ or \sit) kinematics:
the metal-poor and metal-rich LLS branches 
have similar mean velocity widths $\Delta v_{90}$ for \ct\ and \hi. 
Given that metal-poor LLSs are thought to trace galactic inflows and 
metal-rich LLSs trace outflows or recycled material (L13), we find
that the \hi\ line widths are not a good way to distinguish inflowing 
and outflowing gas. Determining the systemic velocities 
of the LLS host galaxies would allow us to look for \hi-galaxy 
velocity offsets, which are predicted by cold-mode inflow models
to be a signature of inflow \citep{St11b}, although the simulations
of \citet{BF14} find that \hi\ radial velocity measurements are generally 
poor at distinguishing inflow from outflow.

If the \hi\ (and \ct) velocity width traces the depth of the potential 
well in which the LLS resides, and hence the mass of the underlying halo
(as is thought to be the case for DLAs), our results 
indicate that the LLS metallicity is independent of the mass.
Alternatively, the lack of a metallicity-kinematics correlation 
(Figure 3) may indicate 
that the LLS kinematics \emph{do not} probe the halo mass, 
which would be the case if LLSs are smaller, sub-virial 
structures with randomly-ordered velocity fields
that do not fill the potential well of their host halo.

At high-redshift ($z\!>\!2$), a clear correlation between 
$\Delta v_{90}$(low ion) and metallicity has been observed in DLAs 
and sub-DLAs \citep{Le06, Pr08, DZ09, Ku10, Ne13}.
This has been interpreted as an indication of an underlying 
mass-metallicity relation in the host galaxies \citep{Le06}.
At lower redshift ($z\!<\!1.5$), DLAs show the $\Delta v_{90}$(\hi)--[Z/H] 
correlation, but the data for sub-DLAs is inconclusive \citep{Me09}. 
Our results reinforce the finding that the trend between metallicity and 
velocity width weakens as one descends in \hi\ column, and they extend this 
finding from DLAs and sub-DLAs down to the LLS regime.

The LLS bimodality is reflected to some degree in the \os\ properties, 
although instead of a clean $N$(\os)-[Z/H] 
correlation, we see a lack of low-metallicity LLSs with strong \os;
high-metallicity LLSs can have either strong or weak \os.
The \os\ line width is correlated with metallicity, with three of the 
four broadest (and strongest) systems also happening to be three of the 
four most metal-rich systems in the sample. The \os\ in these systems can 
naturally be interpreted as being produced by galactic outflows, as has been
discussed in detail for the $z$=0.937 absorber toward \object{PG1206+459} 
\citep{Tr11}.

Our results on the H$^+$ column in the \os\ phase of LLSs indicate 
that a substantial mass of highly-ionized gas is present in these absorbers.
Given an \os\ covering fraction of unity out to 150\,kpc around low-$z$
star-forming galaxies \citep{Tu11a}, which we take as indicative
of the physical size of the \os-bearing CGM,
the high-ion \hw\ mass can be calculated 
as $M_{\rm high-ion}$=$\pi r^2 \mu m_{\rm H}N$(high-ion \hw). Using the values of
log\,$N$(high-ion \hw)$\sim$17.6--20 we derived in \S4.3 and a factor 
$\mu$=1.3 to account for the mass in helium and metal atoms, we find 
$M$(high ion)$\sim$10$^{8.5-10.9}(r/150\,{\rm kpc})^2$\msun. 
This is a separate reservoir than the mass traced by the low and intermediate 
ions in LLSs, which is also estimated to be $\sim$10$^{9-11}$\msun\ 
\citep[L13;][Werk et al. 2014, in prep.]{We13}.
We therefore find that including the \os-traced 
phase of LLSs approximately doubles their contribution to the cosmic baryon 
census, which is still incomplete \citep{Br07, Sh12}.

Our claim that the high-ion phase of LLSs is distinct from the low-ion phase
is based on the kinematic results presented in \S4.2, which indicate that 
all but two of the LLSs have kinematic sub-structure, in the form of velocity 
centroid offsets between the strongest \hi\ and \os\ absorption,
multiple components, or both.
This result is valid independent of any ionization 
modeling, and follows merely from inspection of the absorption-line profiles.
The co-existence of high-ion and low-ion absorption at similar velocities 
has been noted before in many LLSs and in \lya\ forest absorbers at lower 
\hi\ column \citep[e.g.][]{Tr08, WS09, Ho09, Ch12}. 

A separate line of argument that LLSs are multi-phase 
(containing regions of differing density and temperature)
is that single-phase photoionization models systematically under-produce 
the observed \os\ column densities
\citep[][L13]{Co08, Le09, Tu11b, Tr11, Ka12, Cr13a} given 
standard models of the incident ionizing radiation field 
\citep[the UV background;][and later updates]{HM96}.
This under-production of \os\ by photoionization models
is also the case in the Galactic high-velocity 
clouds \citep[HVCs;][]{Se03, Co05, Fo05, Ga05}, gaseous objects lying in the 
low halo of the Milky Way, many of which have \hi\ columns that classify them
as LLSs \citep{Fo06}, although we note that HVCs on average trace gas 
at lower impact parameters than LLSs \citep{LH11, Le12}.
In HVCs the favored mechanism for the production of the high ions
is in the conductive \citep{Gn10}, turbulent \citep{Kw11}, or shocked
\citep{Ha87,DS96} boundary layers 
between the clouds and a surrounding million-degree corona, because these 
models can reproduce the kinematics and line strengths of the HVC high-ion 
absorption profiles, although no single ionization model is able to account
for all the high-ion profiles observed in Milky Way halo directions 
\citep{Wa12, Ma13}.

We stress that in the interface picture, the high ions do not directly 
trace a diffuse 10$^6$\,K corona, but rather the boundary layers where 
such a corona interacts with embedded cool clouds.
The \os\ column observed in a sightline through a single interface
is independent of the cloud's metallicity \citep{He02}, because of 
the physics of radiatively cooling gas: lowering the oxygen 
abundance lengthens the cooling time, so oxygen atoms stay in the \os\ phase 
for longer, compensating for their lower abundance. In this picture, 
the variable that leads to the higher \os\ columns is a larger number of 
components along the line-of-sight.

Alternative models place the \os\ in a diffuse, extended hot halo 
\citep{MM96, Ma03, St12, Sh13, BF13, Hu13}. 
In simulations, cold accretion
(which we argue can be seen in the form of low-metallicity LLSs) 
dominates in low-mass halos, which do 
not possess a million-degree halo since their virial temperature 
is too low. The \os\ absorption in such systems arises from winds,
mostly on their way out, since there is not much wind 
re-accretion in low-mass halos \citep{Op10}.
On the other hand, high-metallicity LLSs are thought to arise in both
low-mass and high-mass halos. In the low-mass case,
the \os\ would arise from outflows, as for low-metallicity LLSs.
In the high-mass case, the \os\ could arise from either inflowing
or outflowing gas, since high-mass halos play host to more 
wind re-accretion \citep{Op10}, or from quasi-hydrostatic halo gas that
is not participating in inflow or outflow.

We conclude by mentioning the lack of \os\ absorption in five of the \nos\
LLSs in our sample, three of which have sensitive (constraining) 
non-detections with log\,$N$(\os)$<$13.6.
One interpretation of these absorbers is that they trace 
the CGM of quiescent, red-and-dead galaxies, either with halos 
too hot for \os\ (although this would leave unanswered the question
of why such halos frequently show \hi),
or with no hot halos at all, as opposed to the CGM of SFGs.
This would be broadly consistent with recent results 
on quiescent galaxies, which generally do not show \os\ absorption 
\citep{Tu11a} but in $\approx$40--50\% of cases
show strong \hi\ [log\,$N$(\hi)$\ga$16] within 150\,kpc \citep{Th12}.
The similar range of \os\ column detected in our LLS sample
and in the COS-Halos SFG sample supports (but does not require)
this interpretation. 
However, two of the sensitive \os\ non-detections are in metal-poor LLSs,
which are expected theoretically to trace low-mass halos, not the halos 
of massive ellipticals, and hence these two require an alternative explanation.

Follow-up imaging of the host galaxies of the full sample of LLSs discussed 
here will allow us to relate the \os\ properties to the galaxy properties.
In particular, such imaging would allow us to compare the \os\  
with the orientation of the host galaxy disk, and hence
look for evidence that \os-bearing outflows 
preferentially occur along the minor axis of the galaxy,
as is the case for \mgw\ absorbers \citep{Ka11, Bou12, Bor13}.
It would also allow us to investigate whether $r$=150\,kpc is 
a suitable impact parameter to use for LLS host galaxies; 
indeed, 11 of the \nos\ LLSs already have a candidate host galaxy
identified within $\approx$120\,kpc (L13).

\section{Summary}
We have studied the high-ionization phase and kinematics
of the circumgalactic medium (CGM) 
around low-$z$ galaxies by surveying \os\ absorption in a sample of \nos\ 
low-$z$ ($0.08\!<\!z\!<\!0.93$) Lyman Limit Systems (LLSs) observed with 
\emph{HST}/COS. 
The data are supplemented in eleven cases by Keck/HIRES spectroscopy,
and in two cases by Magellan/MagE spectroscopy, to provide
coverage of \mgw.
This has allowed us to place empirical constraints on the 
properties of the high-ion phase of the CGM and its relation to 
the low-ion phase, and to search for bimodality in other LLS properties
following the finding of L13 that LLSs have a bimodal metallicity 
distribution. Our key results are as follows:

\begin{enumerate}
\item
\ndetword\ of the \nos\ LLSs show \os\ detections.
Among these, the mean log\,$N$(\os) is 14.22$\pm$0.44. 
We find a moderate trend (97.6\% confidence) in which metal-rich 
([Z/H]$>$--1) LLSs show higher \os\ columns than metal-poor 
([Z/H]$<$--1) LLSs. Metal-poor LLSs only show weak \os; metal-strong LLSs 
can show strong or weak \os. There is no correlation between the LLS \os\ 
column and \hi\ column. 

\item
Among the \ndet\ systems with \os\ detections, all but two show
evidence of kinematic sub-structure, in the form 
of velocity centroid offsets between the strongest components of 
\os\ and \hi\ absorption of more than 10\kms, multiple components, or both. 
The centroid offsets extend up to $\pm$50\kms\ around the systemic 
redshift, and are not a function of LLS metallicity. 
The mean absolute centroid offset from \hi\
increases from \cw\ to \ct\ to \os:
$\langle|\delta v_0|\rangle$(\cw--\hi)=6$\pm$1\kms, 
$\langle|\delta v_0|\rangle$(\ct--\hi)=10$\pm$2\kms, and
$\langle|\delta v_0|\rangle$(\os--\hi)=14$\pm$4\kms. 
Thus we find a tentative trend in which
the higher the ionization state, the larger the average velocity
offset from the \hi\ of the strongest absorption component.

\item 
We compare the measured velocity widths $\Delta v_{90}$ for \hi, 
\cw, \mgw, \ct, and \os\ with metallicity. We find no significant 
difference in the distribution of $\Delta v_{90}$(\ct) and 
$\Delta v_{90}$(\hi) between the low-metallicity and high-metallicity 
sub-samples. \emph{Thus the bimodality of LLSs in their 
metallicities is not reflected in their \hi\ kinematics or their
intermediate-ion kinematics.} 
However, for \os, there is a correlation between 
velocity width and metallicity significant at the 99.0\% level. 
The mean observed velocity widths
$\langle\Delta v_{90}\rangle$ are 98$\pm$12\kms\ for \hi,
65$\pm$8\kms\ for \cw, 41$\pm$8\kms\ for \mgw, 
83$\pm$14\kms\ for \ct, and 97$\pm$10\kms\ for \os.
The higher mean line width for \os\ than \cw\ reinforces the
conclusion that LLSs are generally multiphase systems.
The narrowness of the mean \cw\ line width indicates that LLSs
are sub-virial objects.

\item
We calculate the H$^+$ column in the \os-bearing gas, under a range of scenarios
for how the metallicity in the low-ion and high-ion phases are connected,
ranging from the case where the metallicities are the same in each phase
(condensation model) to the case where the high-ion phase has solar 
metallicity (outflow or recycled wind model).
\emph{In any of these three cases}, LLSs contain a significant reservoir of 
highly ionized material, with log\,$N$(high-ion \hw)$\sim$17.6--20,
corresponding to a mass $M$(high-ion)$\sim$10$^{8.5-10.9}$\msun\ 
out to impact parameters of 150\,kpc. This is of similar order to the 
mass contained in the low-ion phase of LLSs \citep[L13;][]{We13}. 
Accounting for the \os\ phase therefore approximately 
doubles the contribution of LLSs to the cosmic baryon budget.
These results show that \emph{all} LLS-traced CGM, whatever kind of 
galaxy lies underneath, contains a significant fraction of baryons 
in its high ion phase.
\end{enumerate}

{\it Acknowledgments}\\
We thank John Stocke and Molly Peeples for helpful comments,
and Tim Heckman and Brice M\'enard for useful discussions.
Support for programs \#11598, \#11741, and \#12854 was provided by 
NASA through grants from the Space Telescope Science Institute, which is 
operated by the Association of Universities for Research in Astronomy, 
Incorporated, under NASA contract NAS5-26555. 
N.K. acknowledges support from NASA grant NNX10AJ95G.

\begin{appendix}

In the first set of figures below,
we present normalized flux profiles of \hi, a low ion (\cw\ or \siw), 
an intermediate ion (\ct\ or \sit), and the high ion \os\ in each LLS.
Where Keck/HIRES or Magellan/MagE data exist, we have added the 
profile of \mgw\ $\lambda$2796 or \mgw\ $\lambda$2803 to the bottom panel.
For \hi, a single Lyman series line (unsaturated where possible) is 
chosen to represent the component structure. 
For the low ions, either \cw\ $\lambda$903.96, \cw\ $\lambda$1036, \cw\ 
$\lambda$1334, \siw\ $\lambda$1020, or \siw\ $\lambda$1260 is shown,
using the line that best displays the component structure
without saturation. For the intermediate ions, the choice of \ct\ 
$\lambda$977 or \sit\ $\lambda$1206 depends on redshift or saturation.
The systems are ordered alphabetically by target name. 
The interval between the two red tick marks in each panel 
indicates the velocity width of absorption $\Delta v_{90}$; this value 
(in km\,s$^{-1}$) is annotated on each panel. Gray shading is used when 
significant absorption ($>$3$\sigma$) is seen in a given line. 
Only unblended \os\ lines are shown (either $\lambda$1031, $\lambda$1037, 
or both). Further spectral data for some (but not all) of these systems 
are given in L13. Note that here we adopt the redshifts given to four 
significant digits in Table 1 of L13; small velocity shifts may be 
observed between these figures and those shown in L13.
This does not affect our kinematic analysis as we are interested in velocity 
differences. Blends caused by the \lya\ forest or other contaminating lines
are identified on the plots.

In the second set of figures, we compare the apparent column density
profiles of \hi, \cw, \ct, and \os\ in each LLS. 
Gaussian fits to the strongest component of absorption are used to determine 
the velocity centroid (relative to the systemic redshift of the absorber) 
of each species. The strongest-component centroid offsets 
$v_0$(\os--\hi), $v_0$(\ct--\hi), and $v_0$(\cw--\hi) 
are annotated on the panels. The scale factors were selected
to normalize the peak of each profile to the same level, for ease of 
comparison.

\clearpage
\begin{figure}\epsscale{1.0}\plottwo{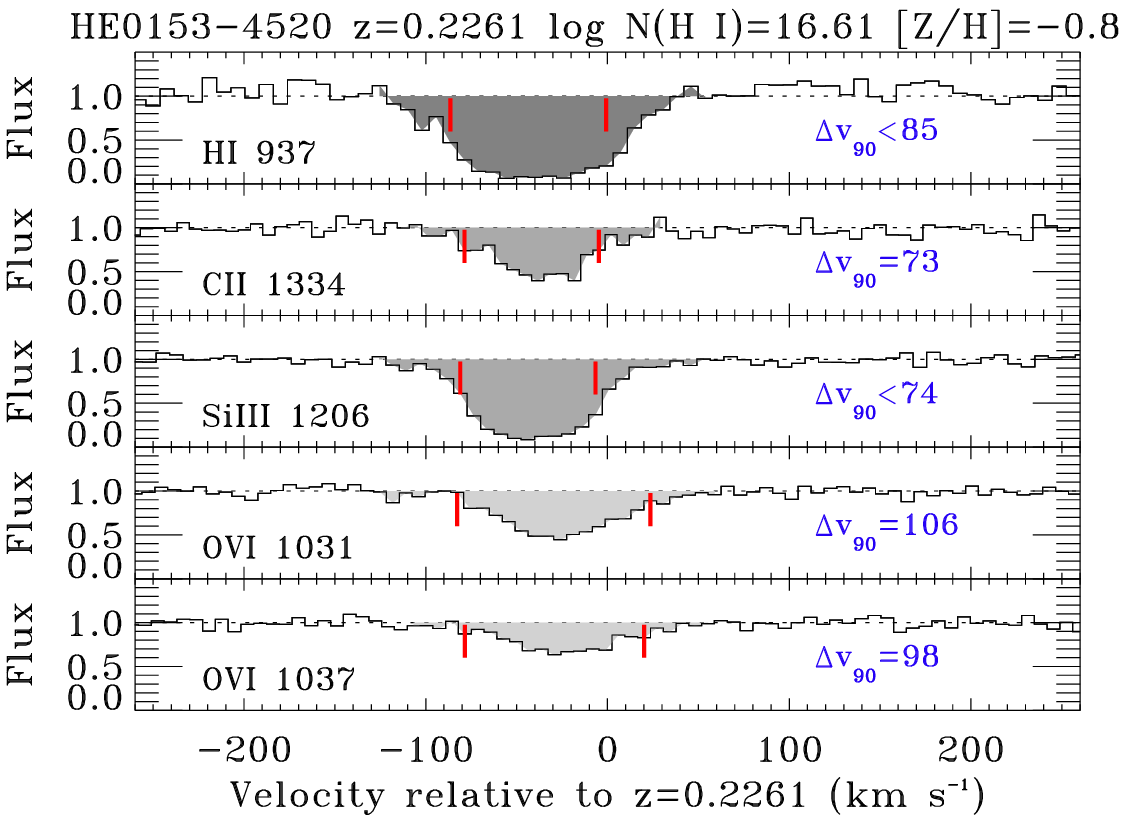}{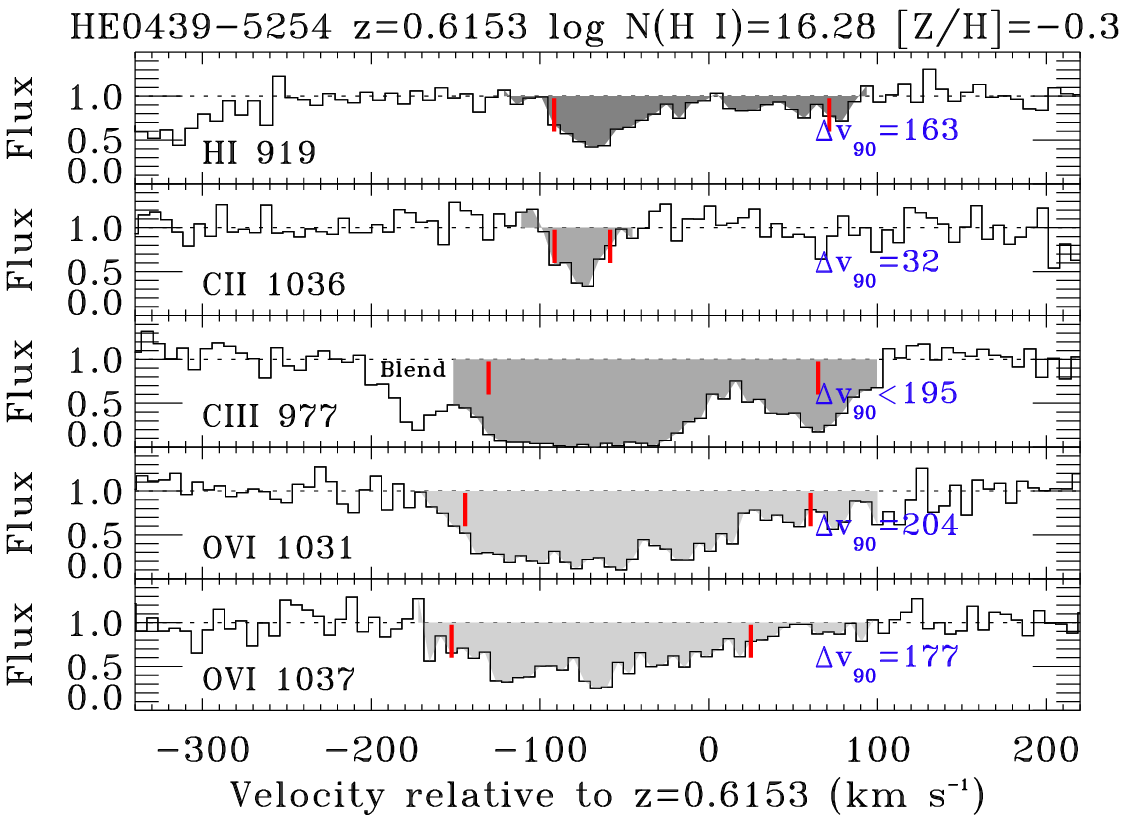}\end{figure}
\begin{figure}\epsscale{1.0}\plottwo{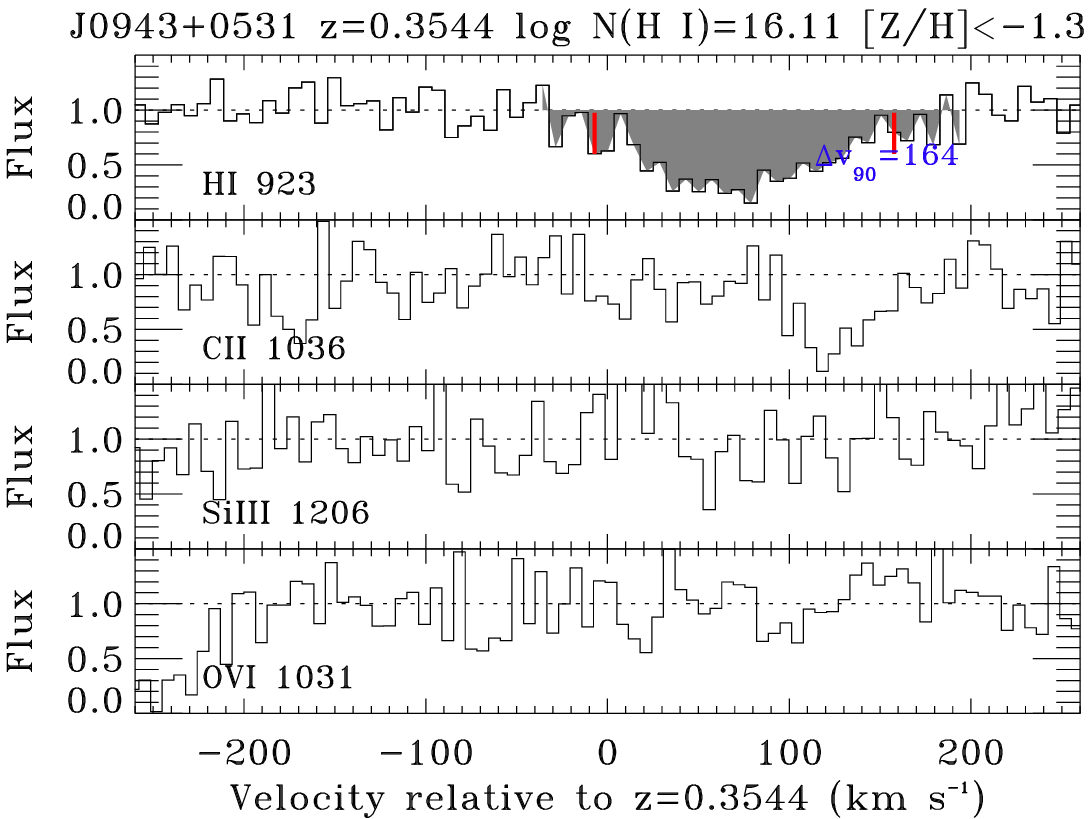}{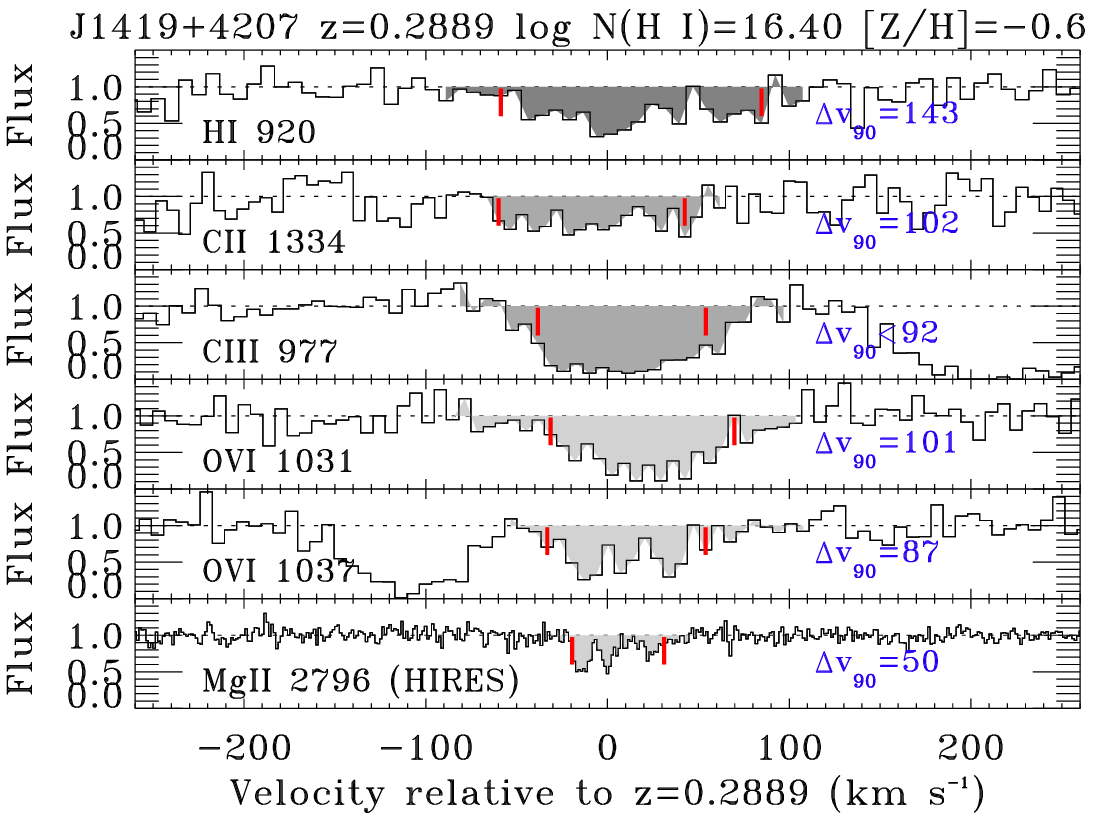}\end{figure}
\begin{figure}\epsscale{1.0}\plottwo{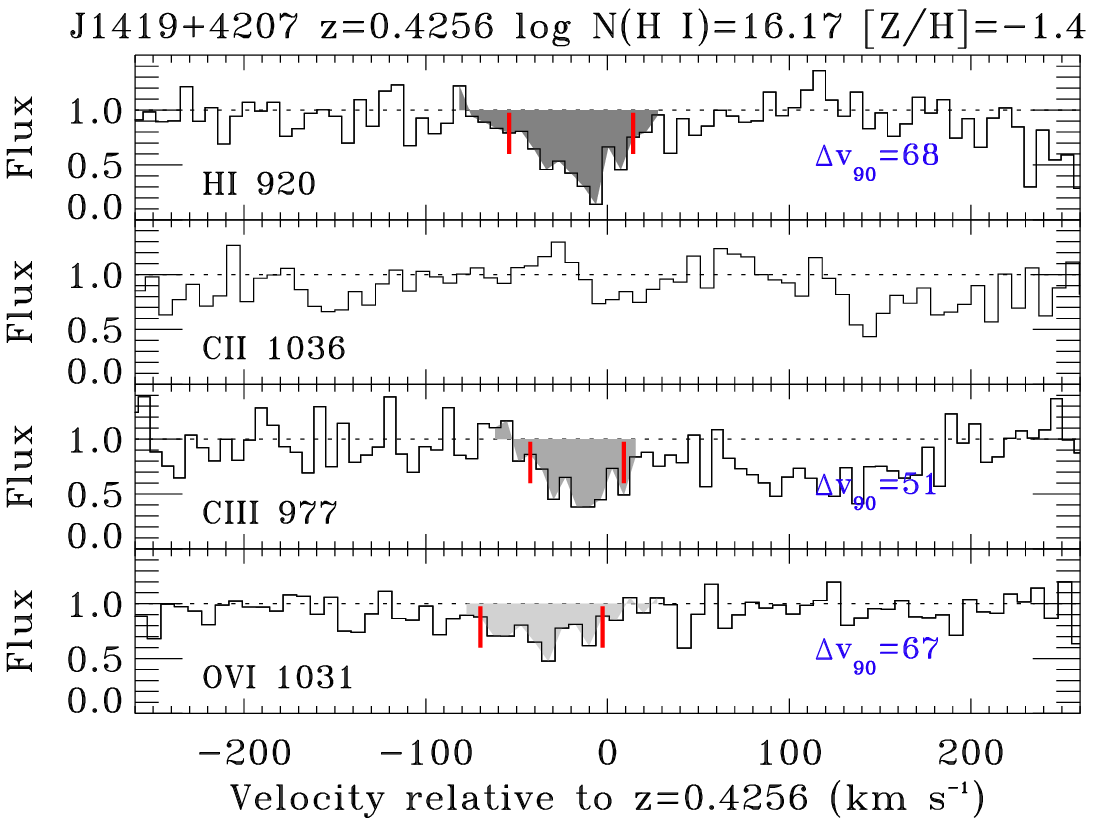}{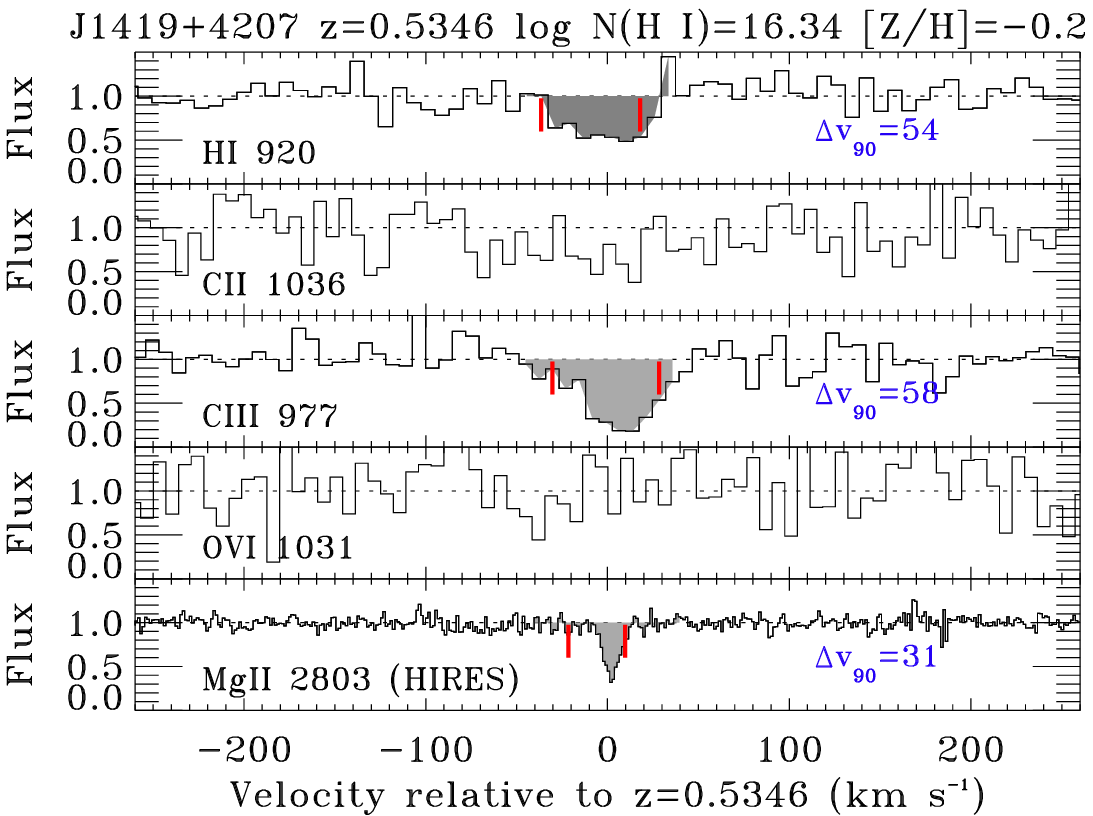}\end{figure}
\begin{figure}\epsscale{1.0}\plottwo{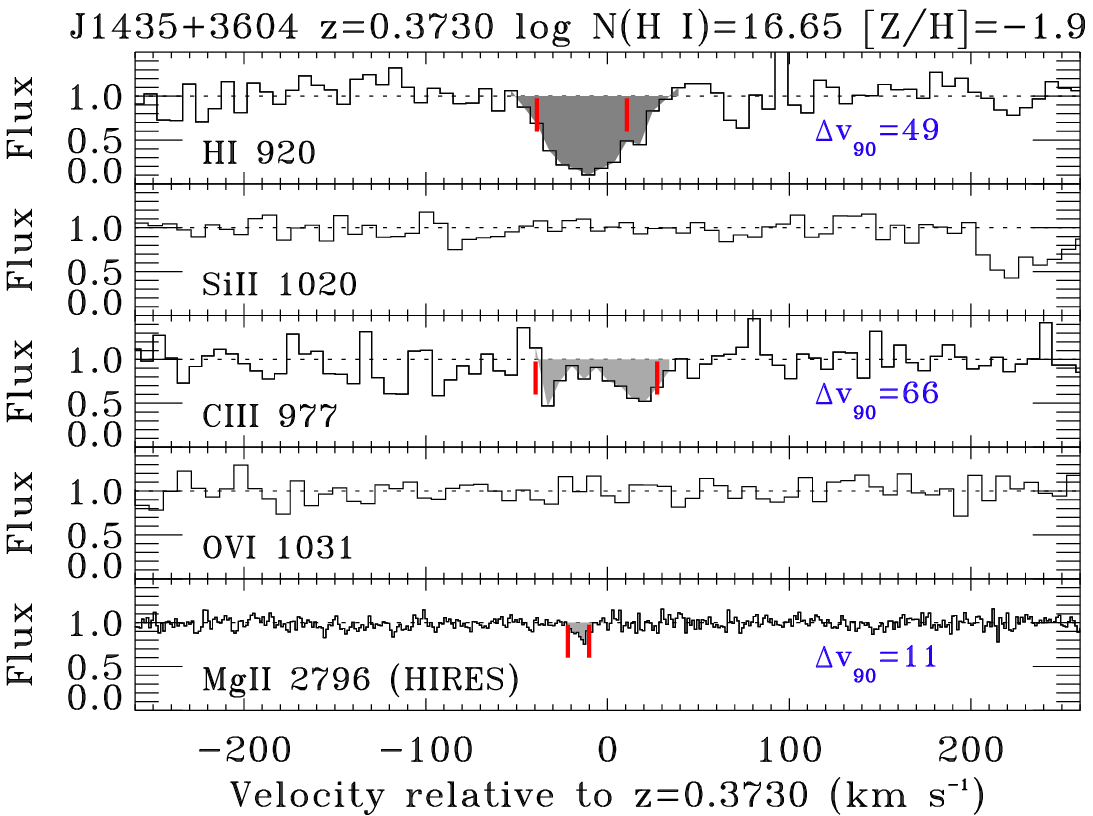}{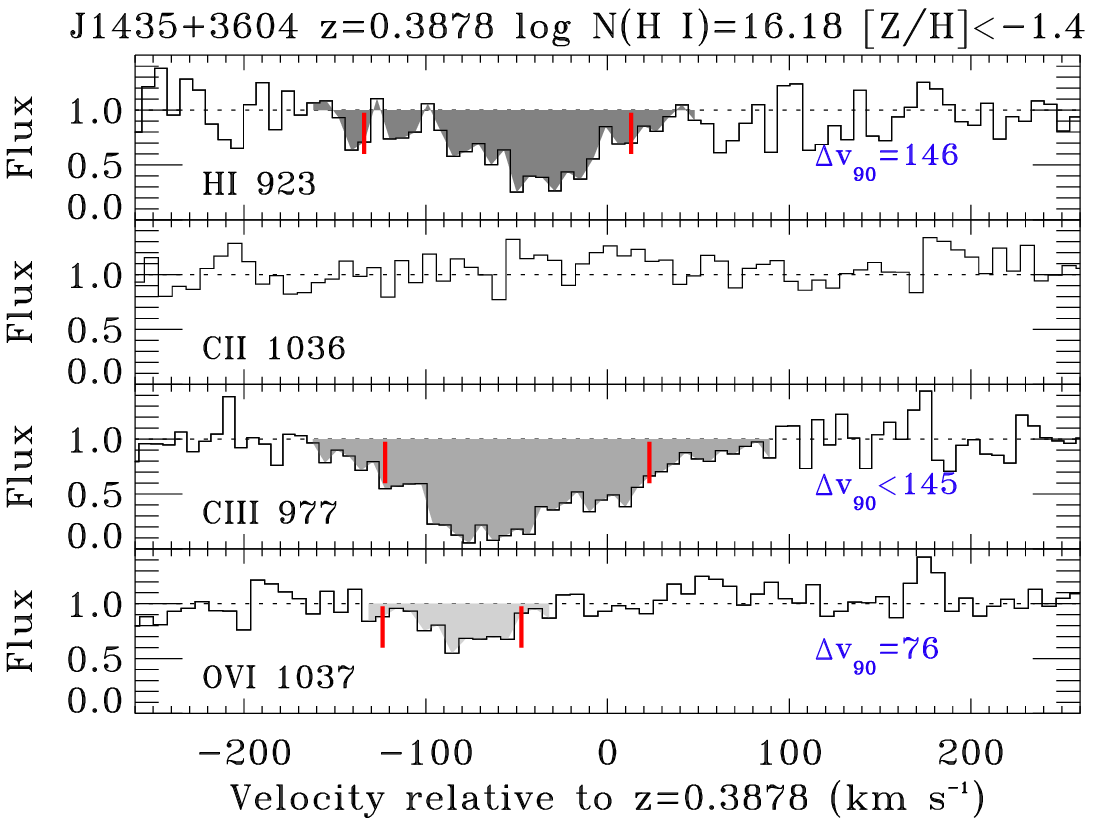}\end{figure}
\begin{figure}\epsscale{1.0}\plottwo{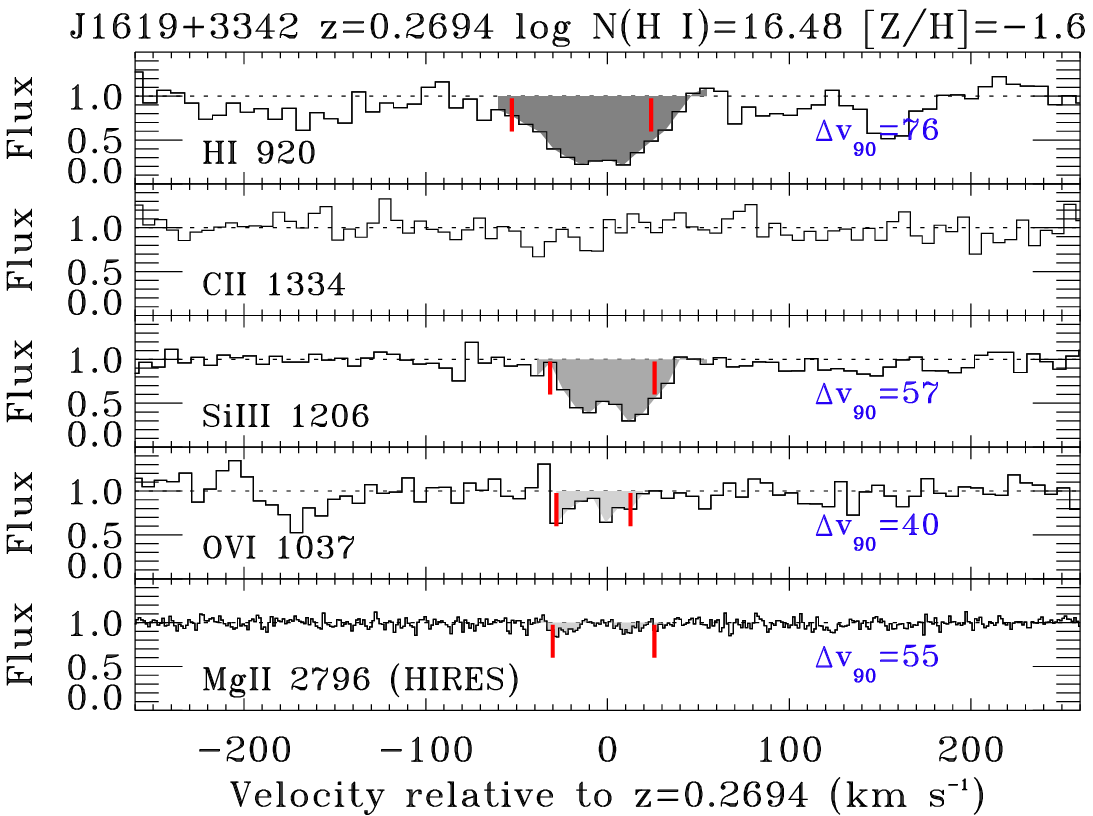}{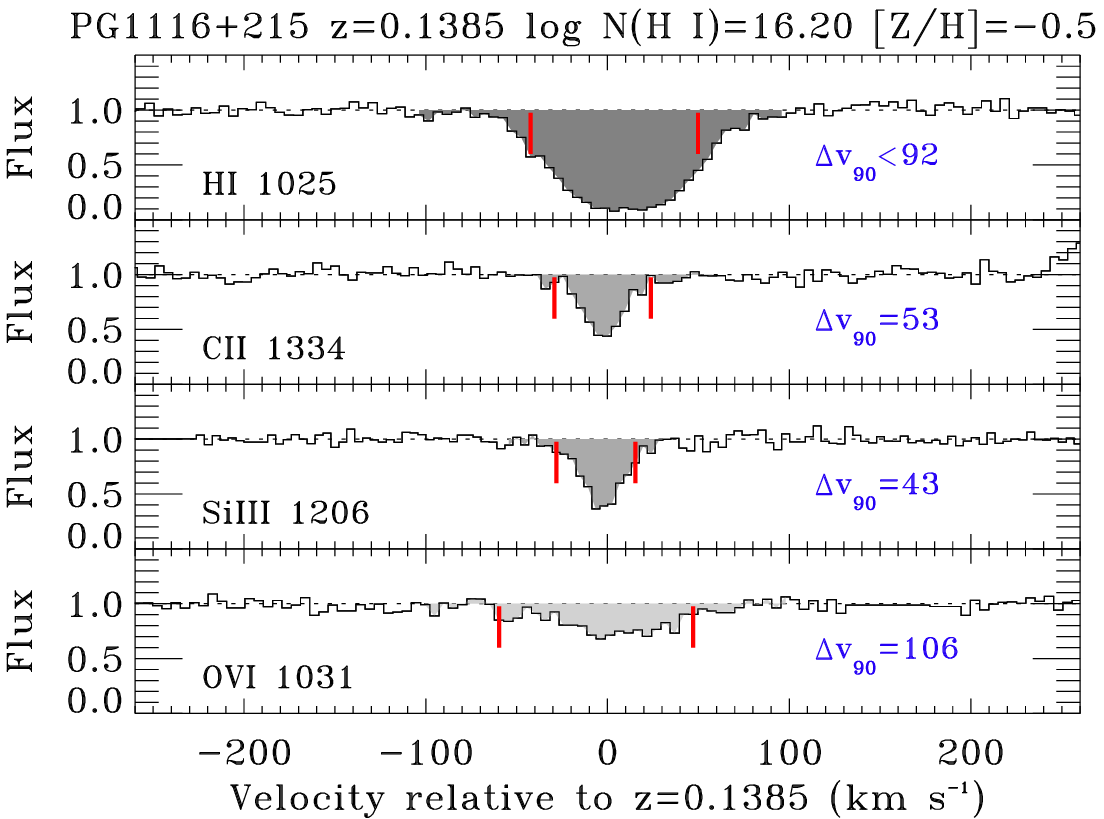}\end{figure}
\begin{figure}\epsscale{1.0}\plottwo{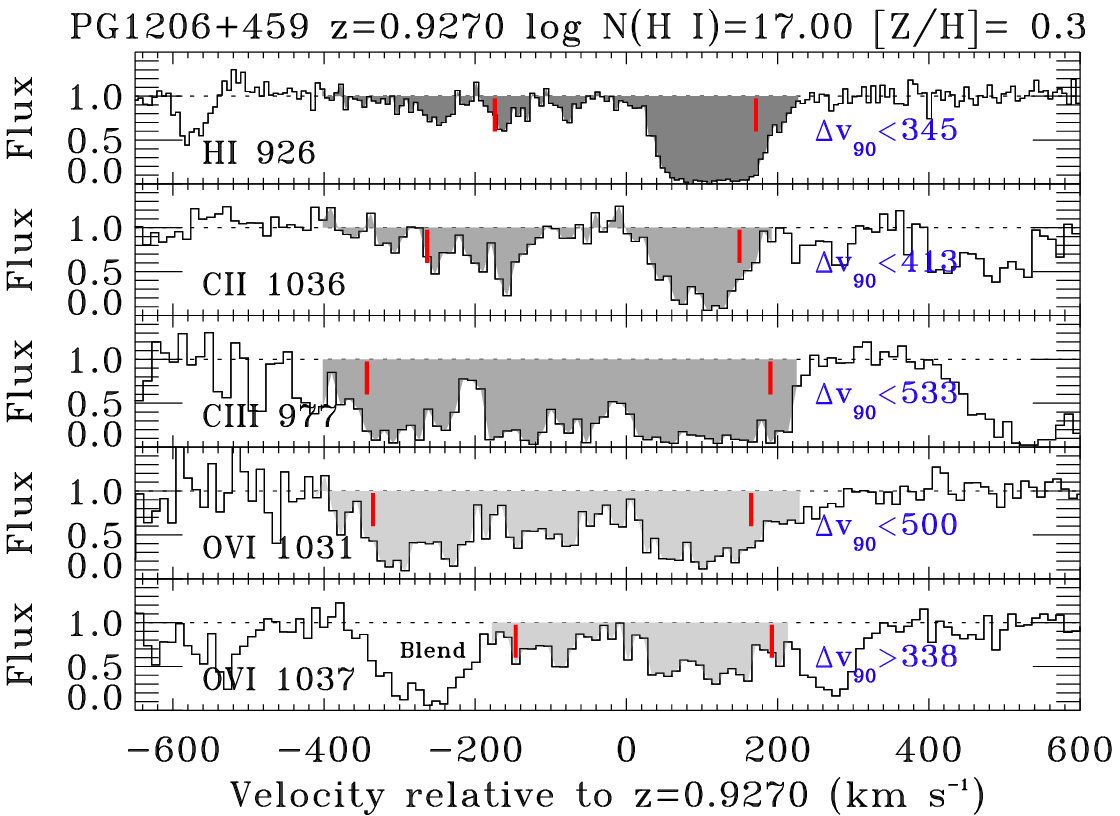}{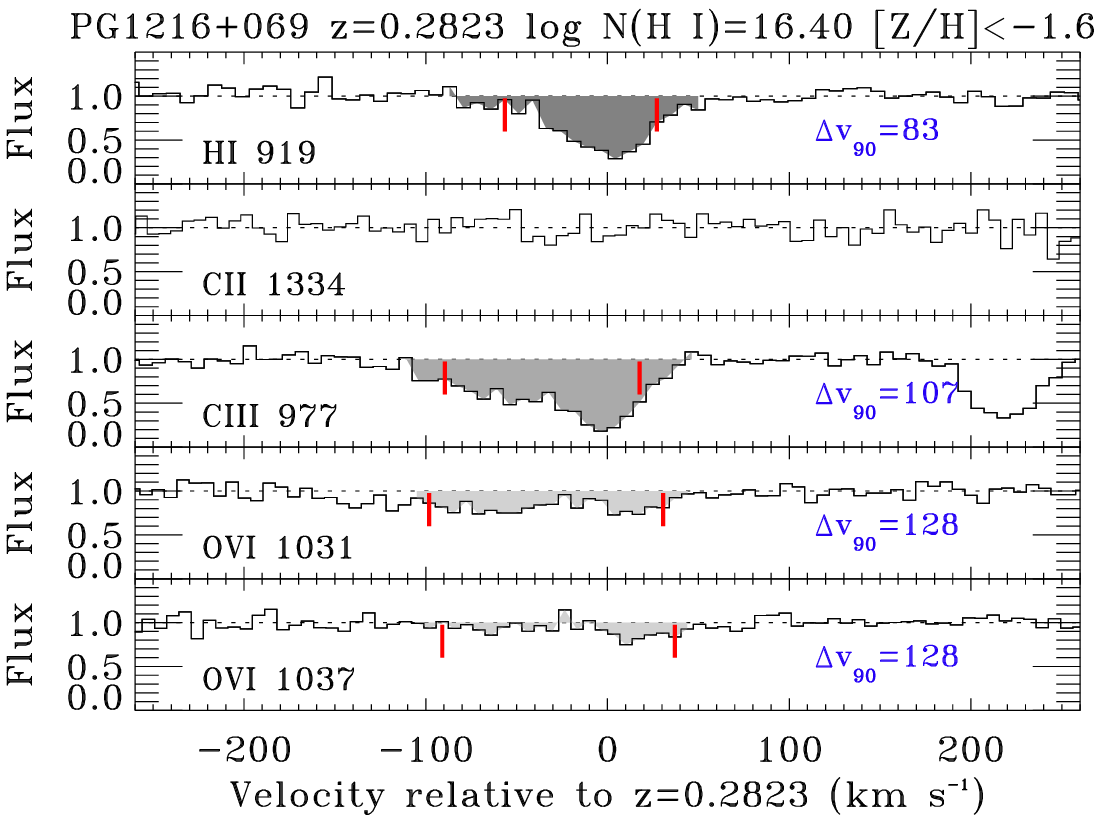}\end{figure}
\begin{figure}\epsscale{1.0}\plottwo{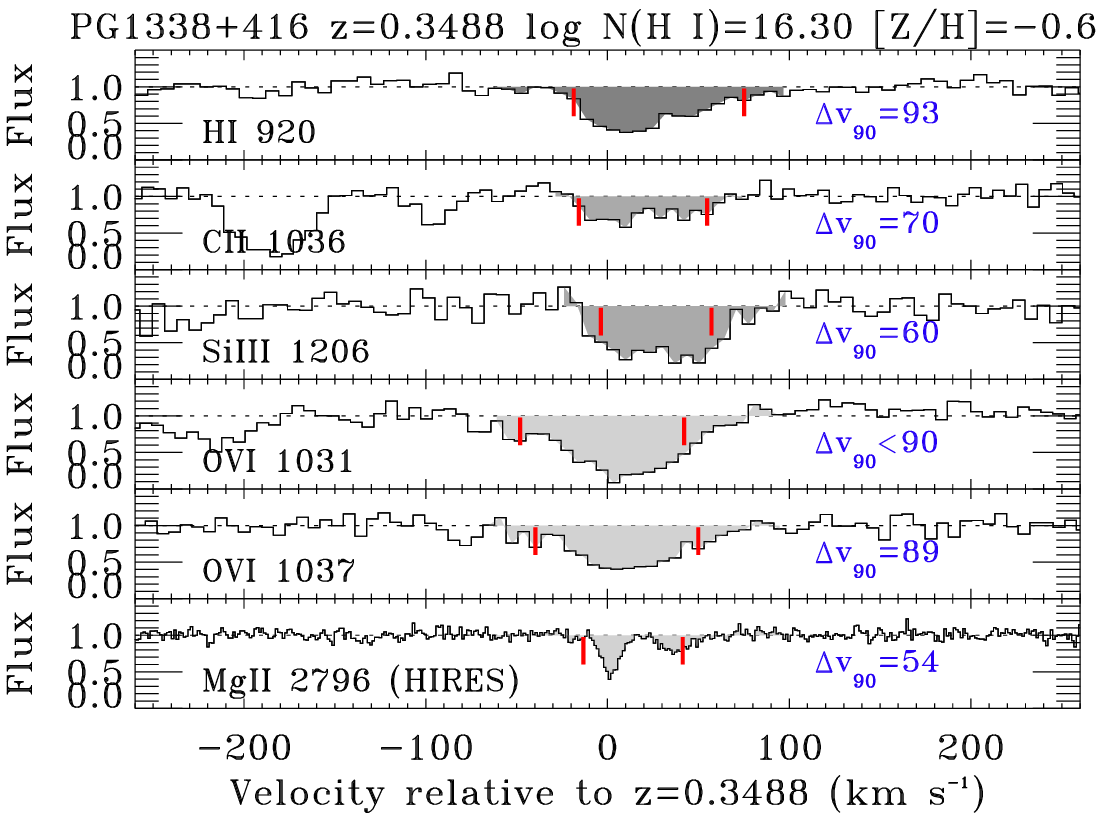}{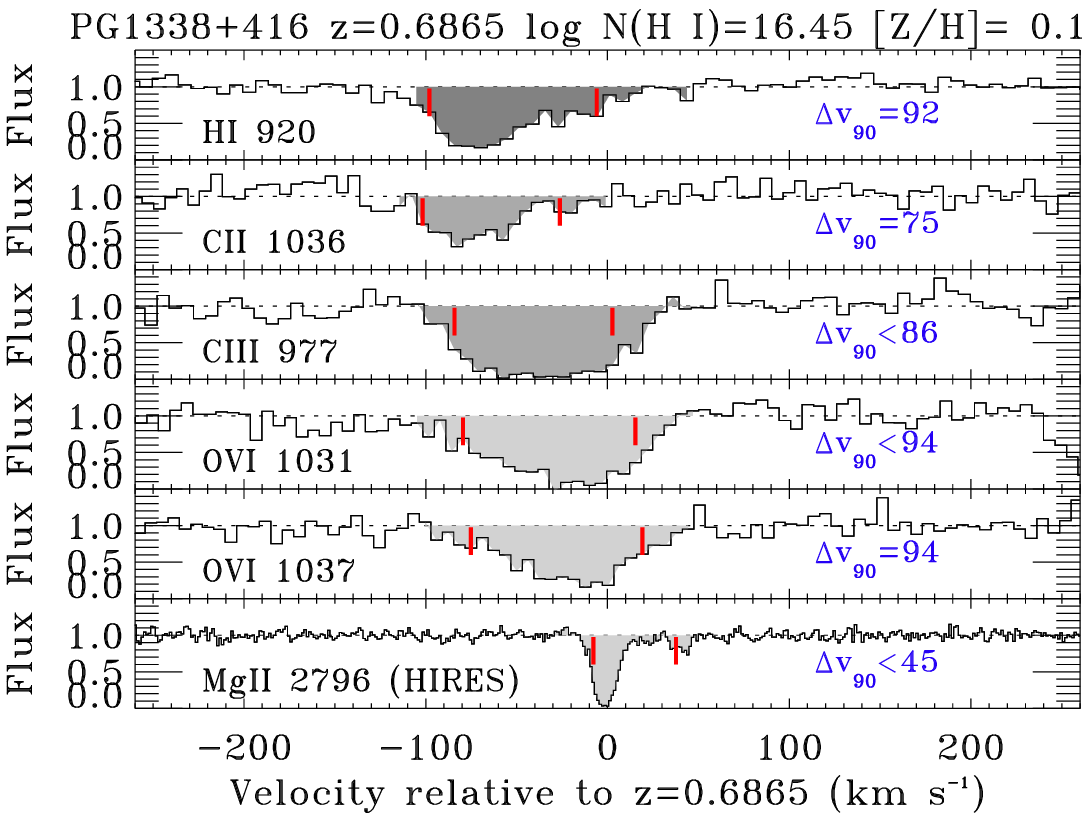}\end{figure}
\begin{figure}\epsscale{1.0}\plottwo{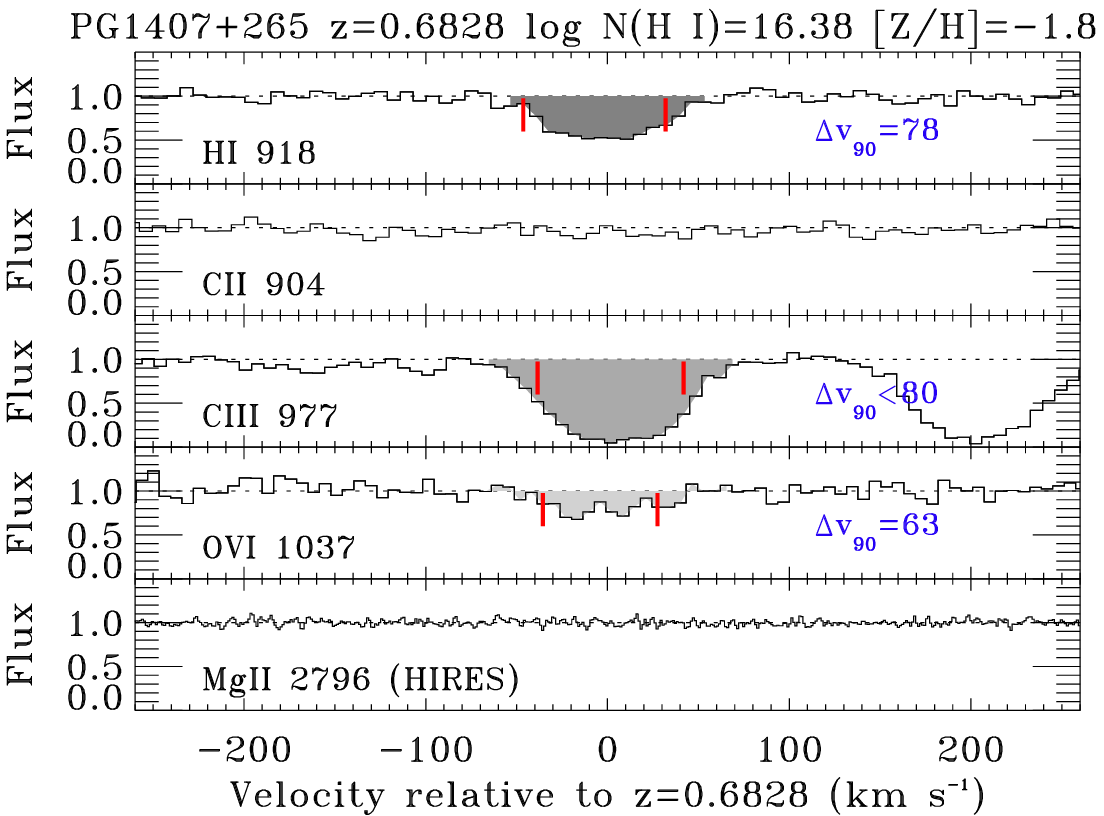}{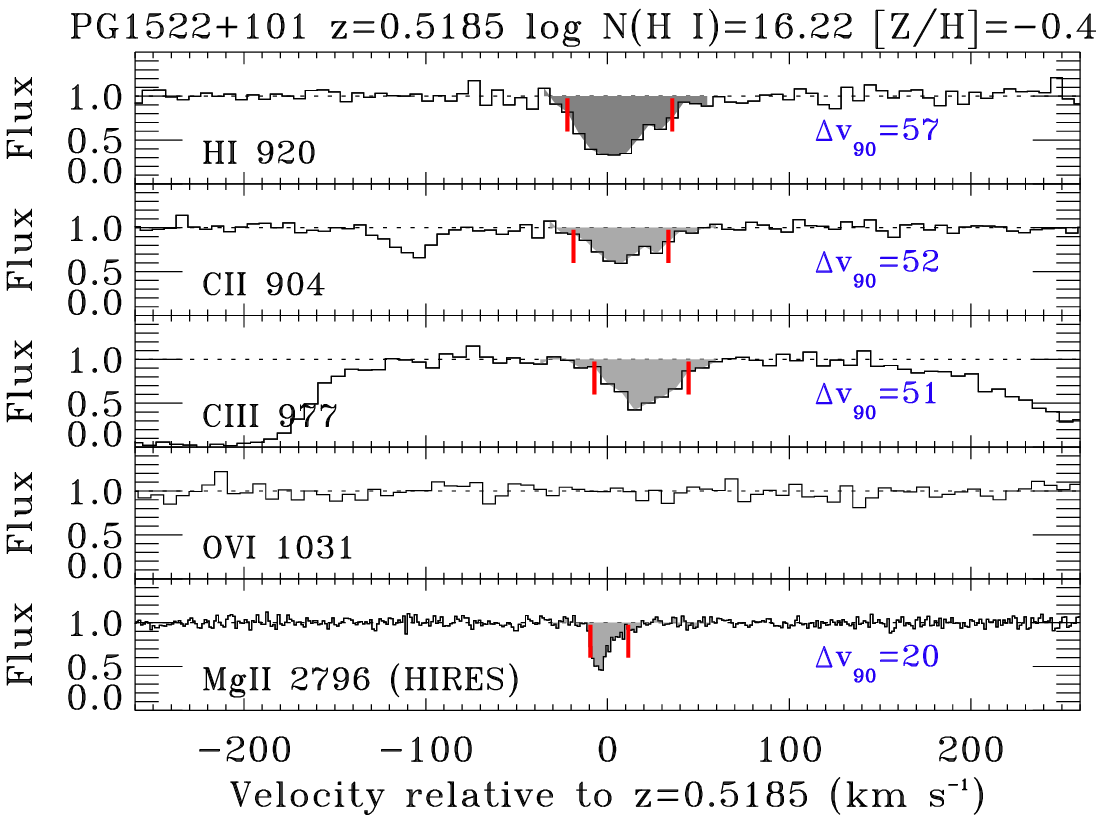}\end{figure}
\begin{figure}\epsscale{1.0}\plottwo{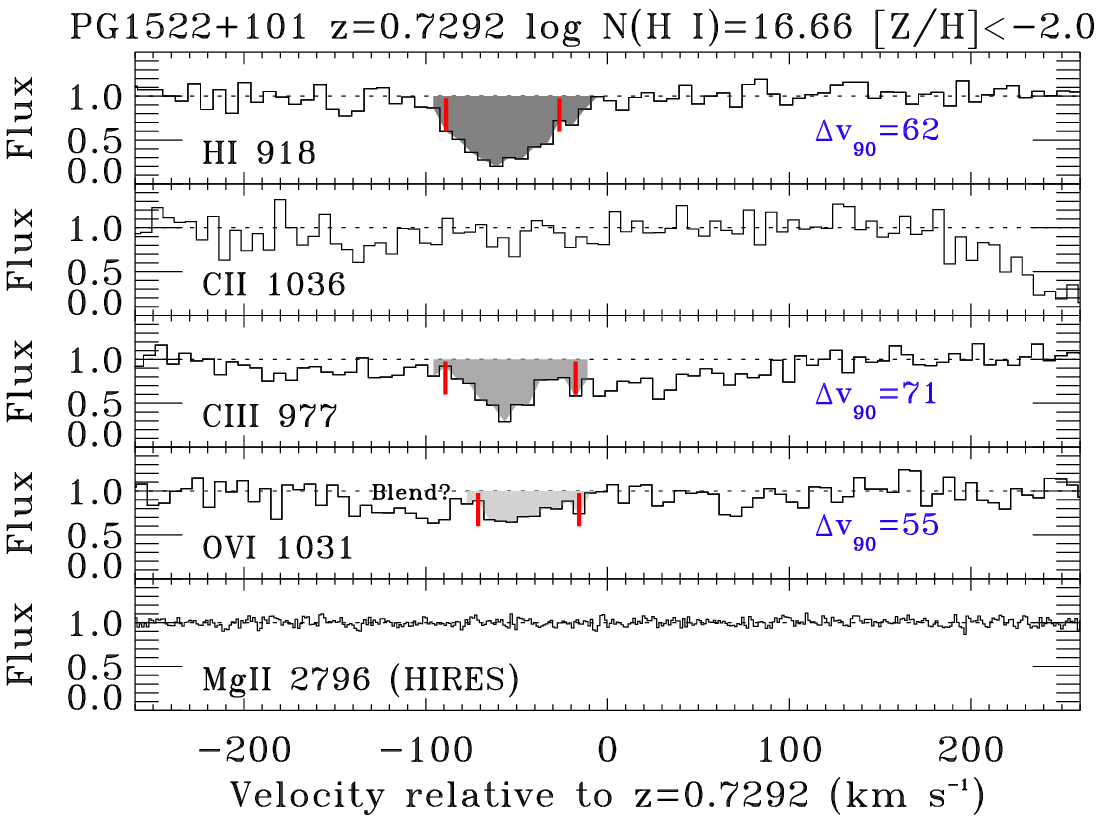}{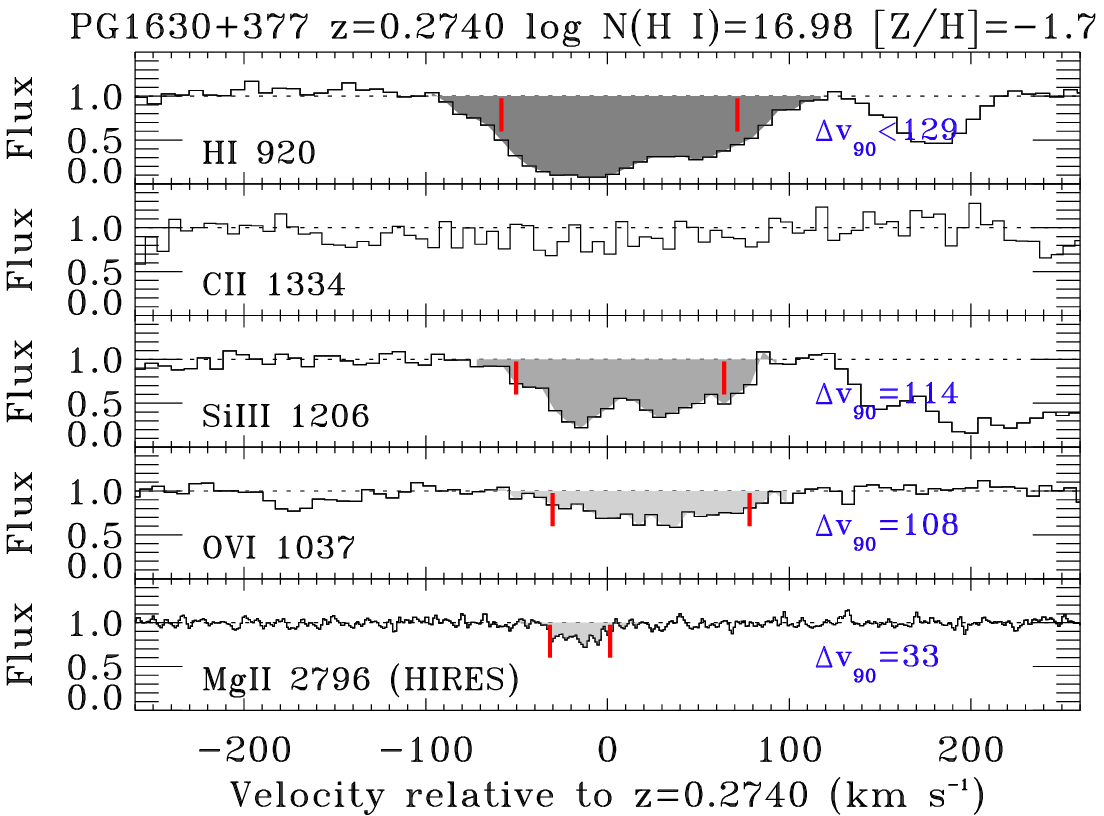}\end{figure}
\begin{figure}\epsscale{1.0}\plottwo{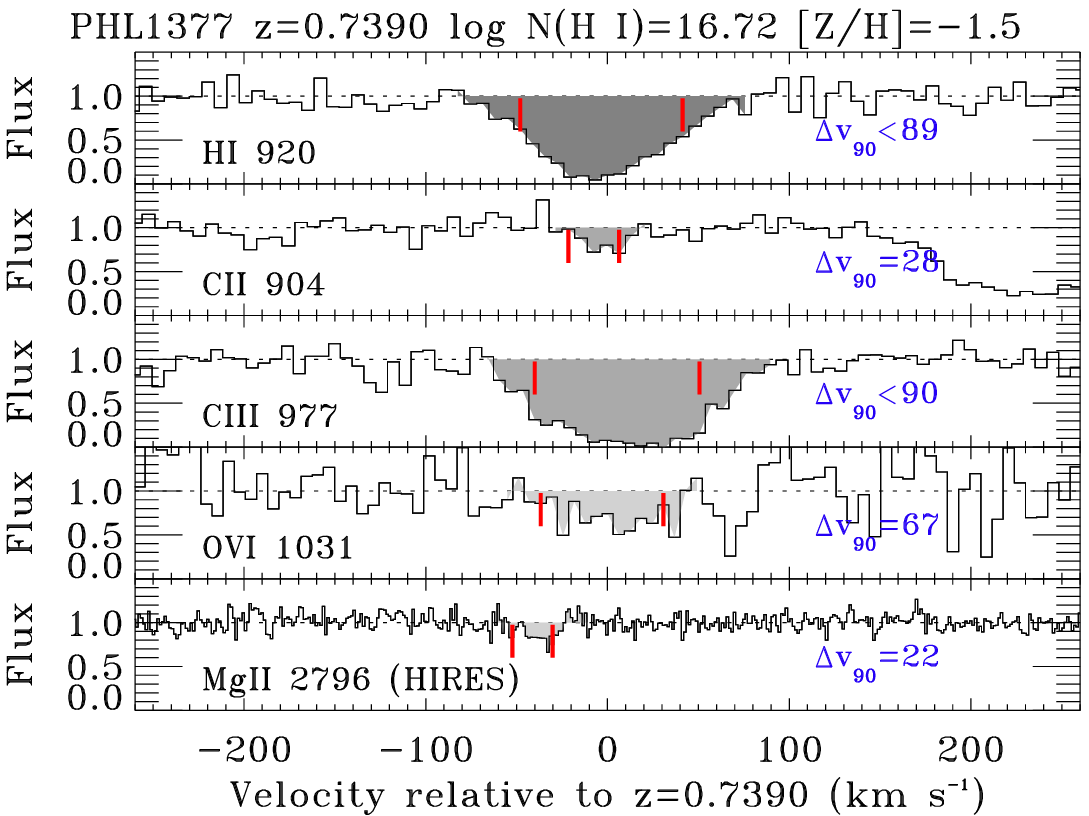}{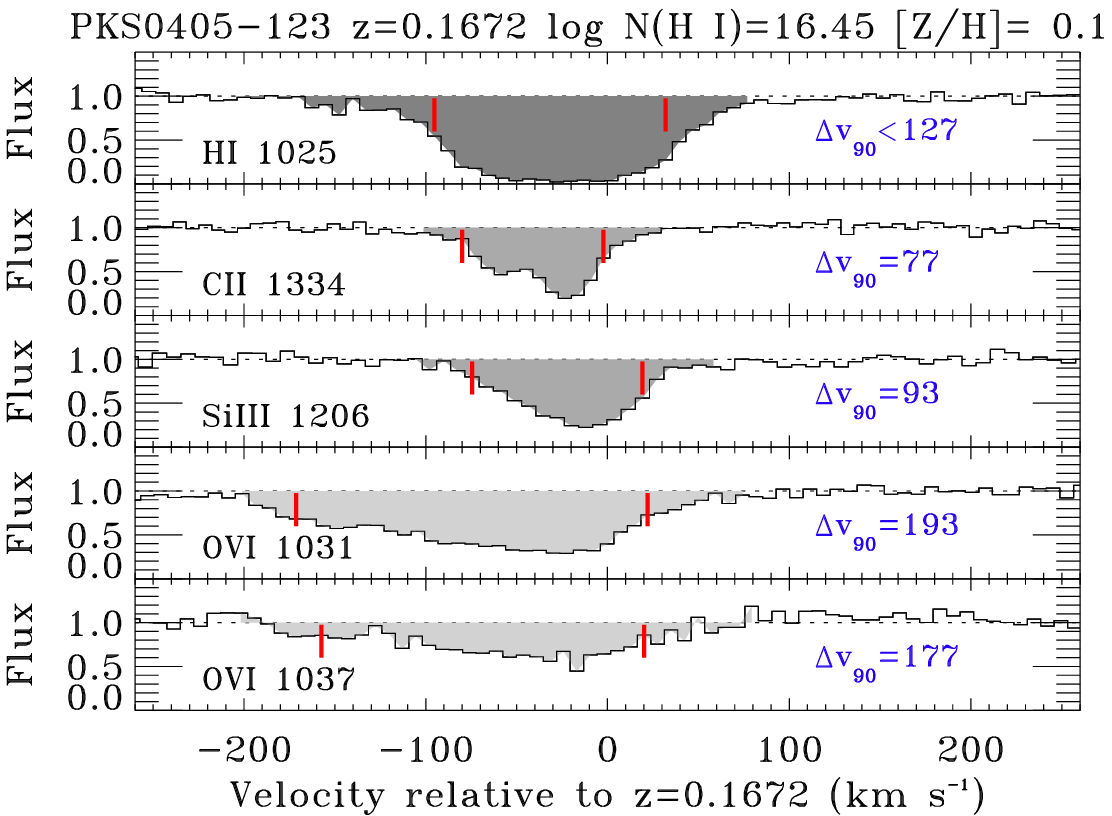}\end{figure}
\begin{figure}\epsscale{1.0}\plottwo{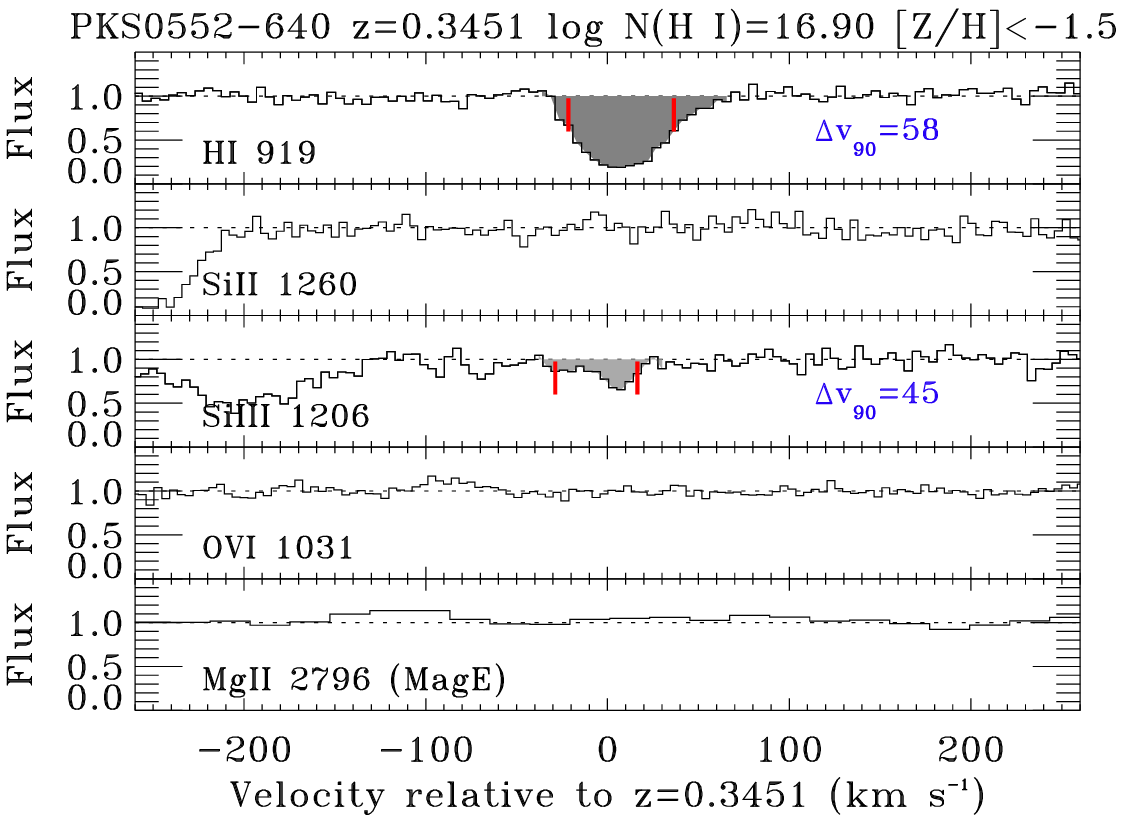}{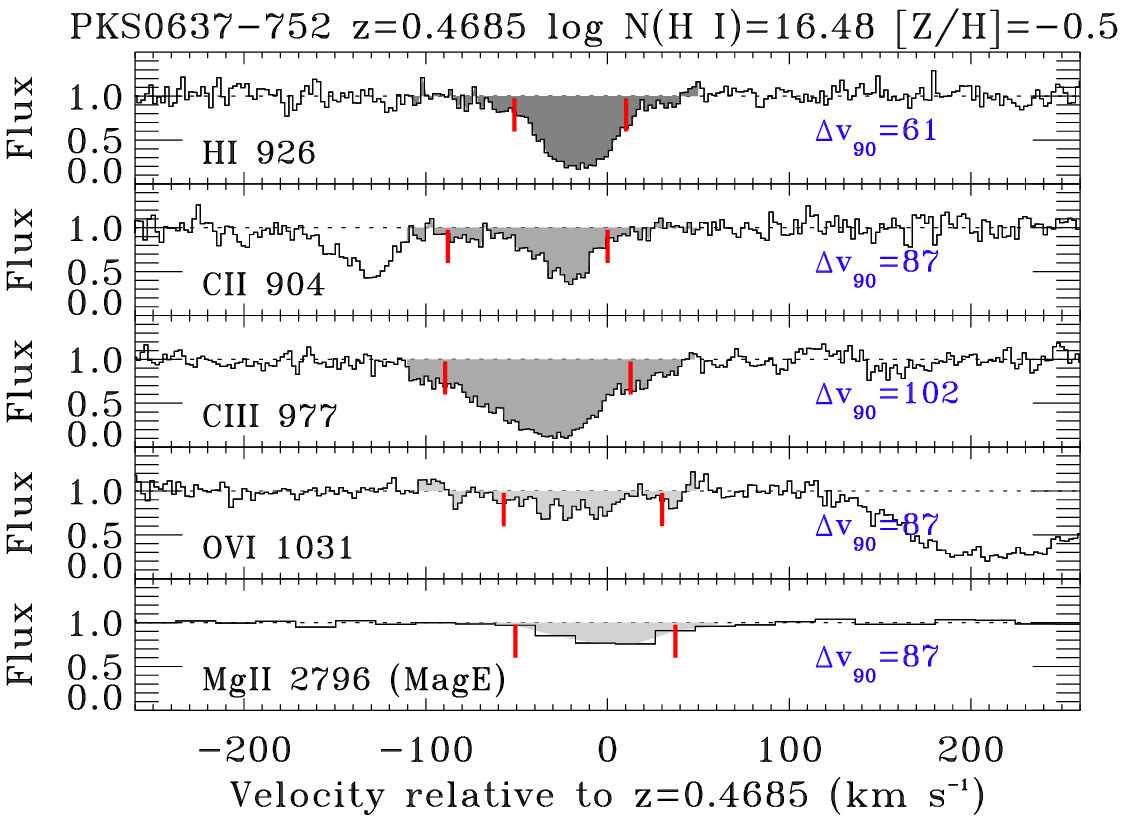}\end{figure}
\begin{figure}\epsscale{1.0}\plottwo{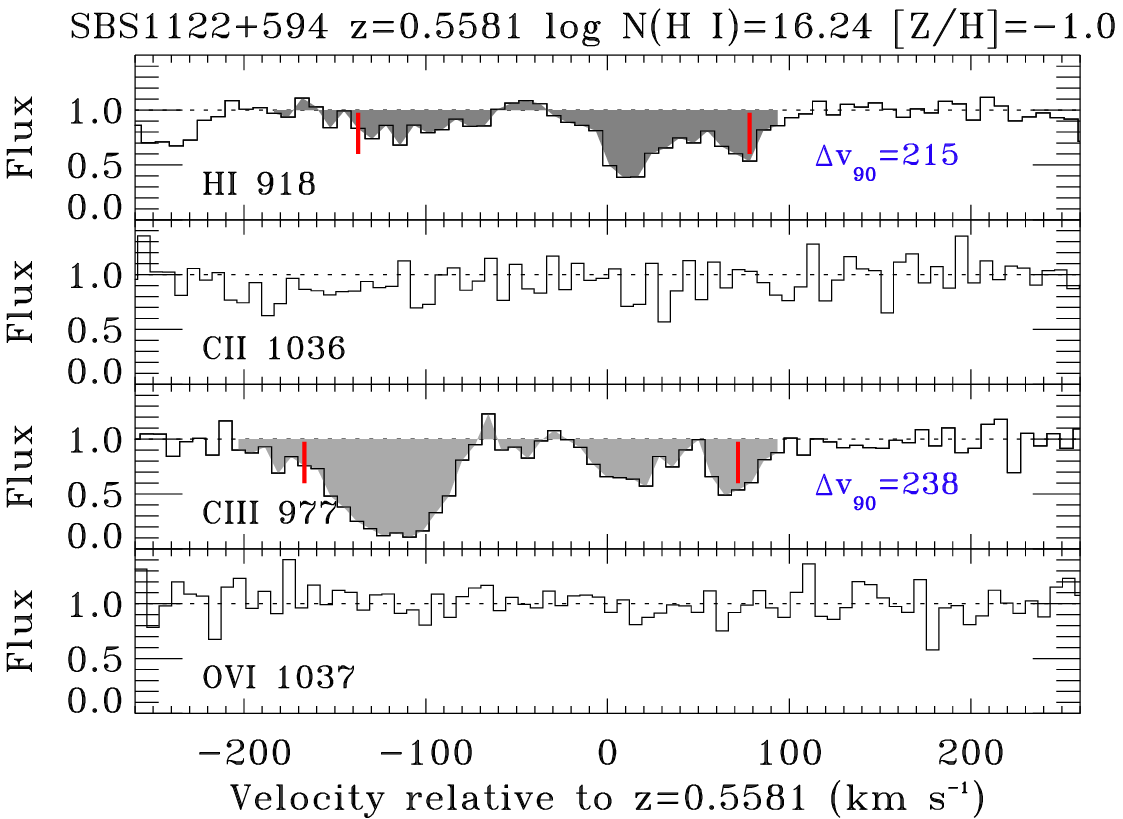}{SBS112.z=0.5581.eps}\end{figure}

\clearpage
\begin{figure}\epsscale{1.0}\plottwo{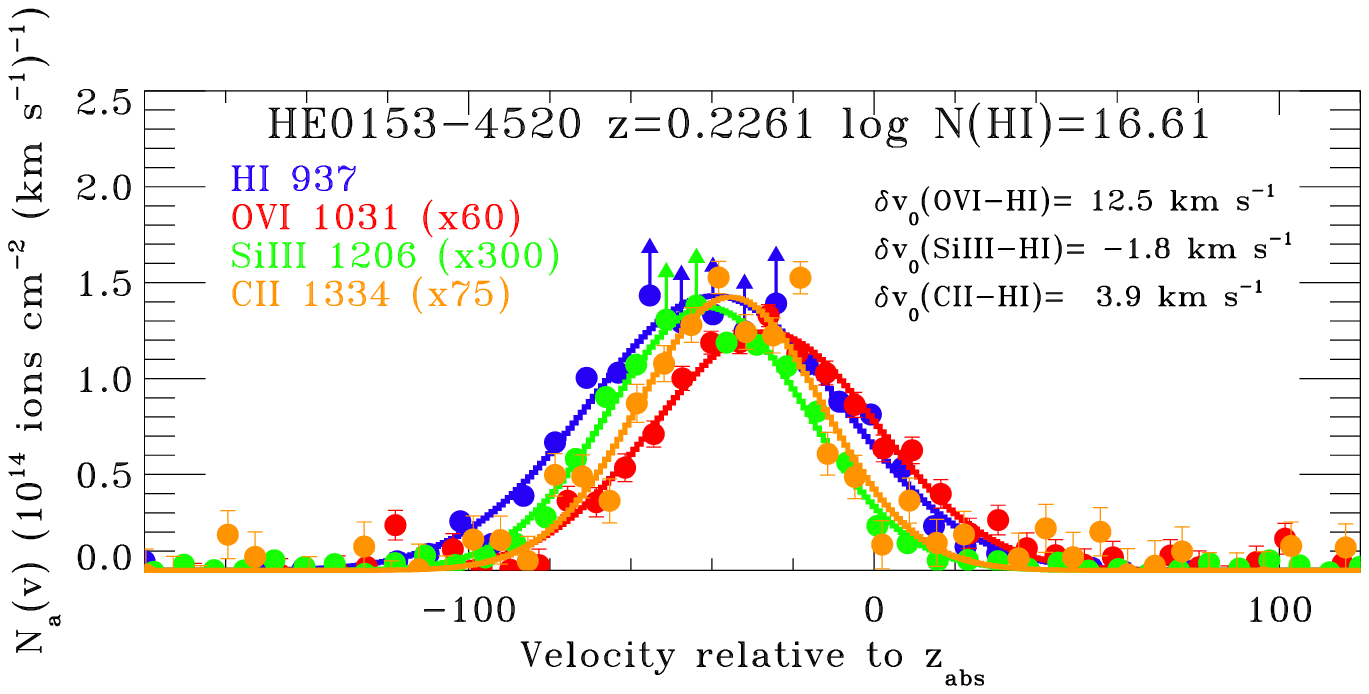}{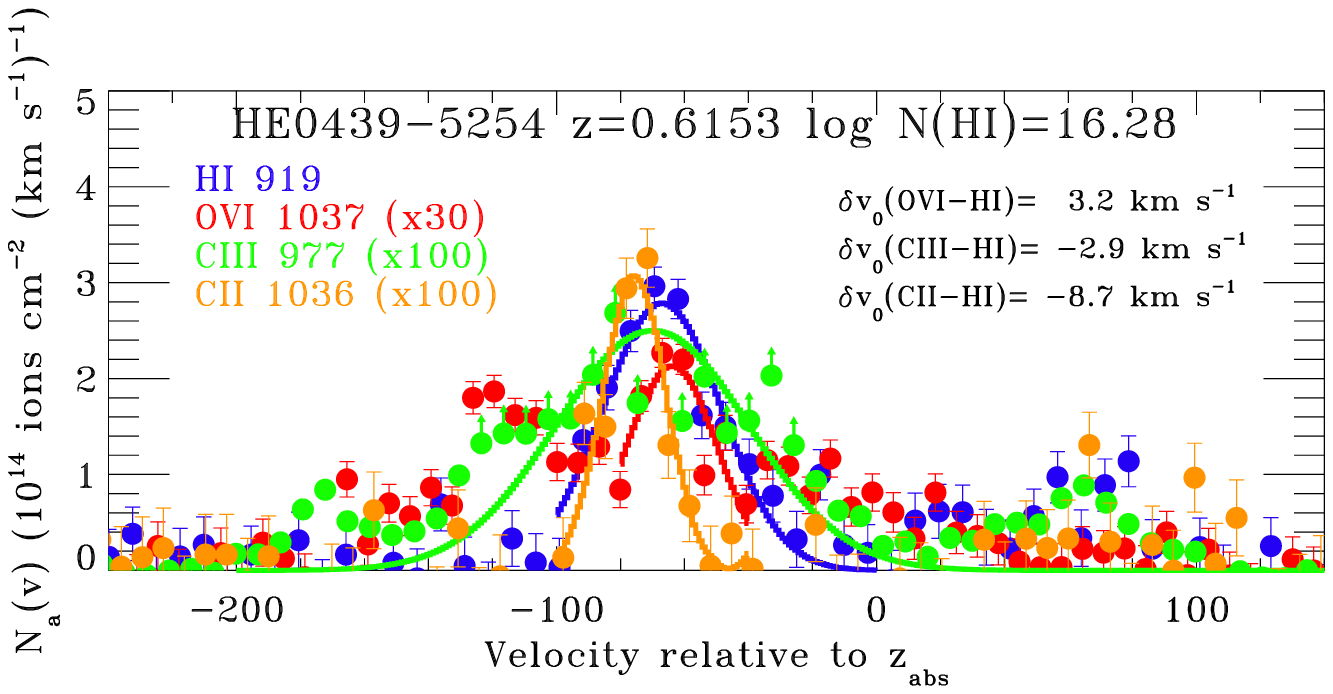}\end{figure}
\begin{figure}\epsscale{1.0}\plottwo{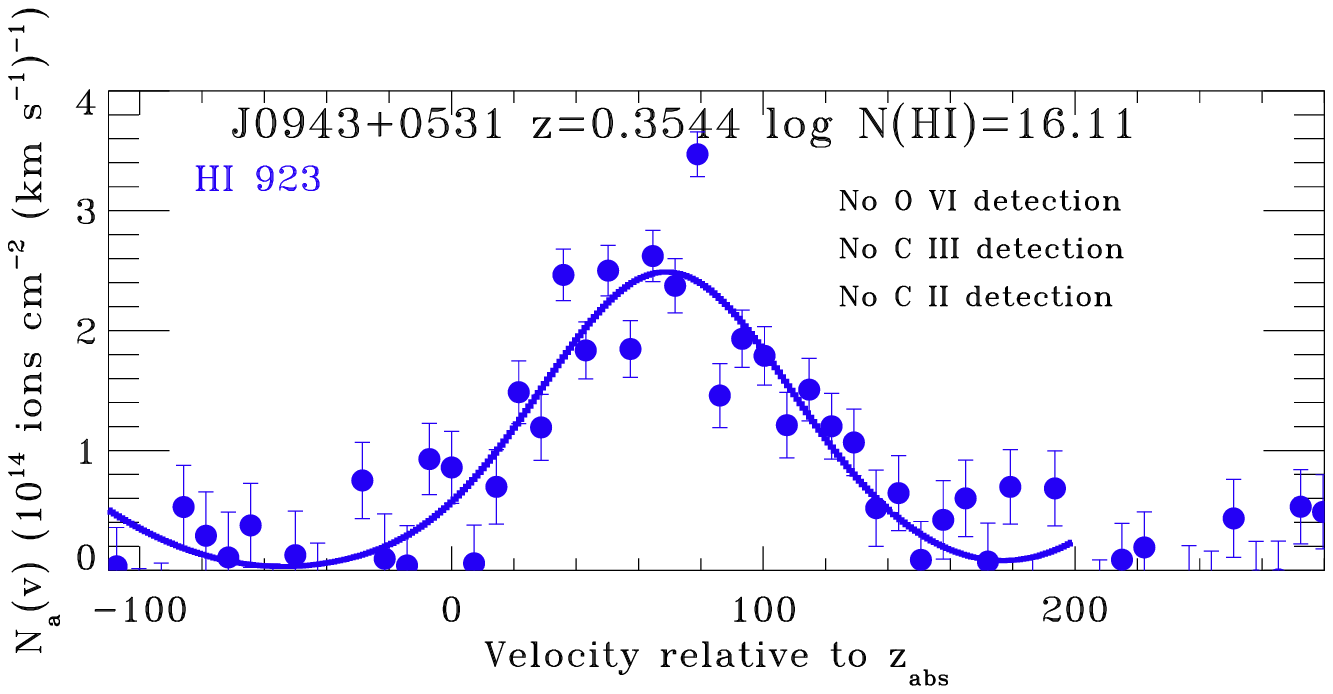}{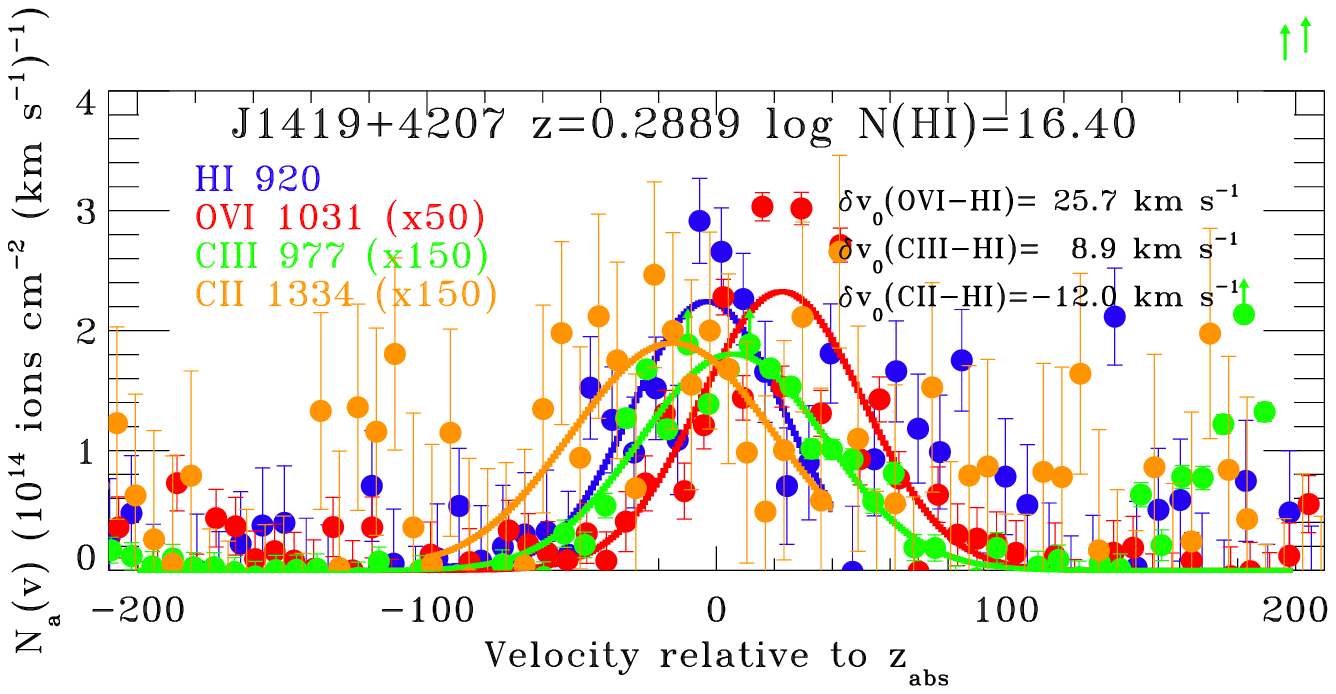}\end{figure}
\begin{figure}\epsscale{1.0}\plottwo{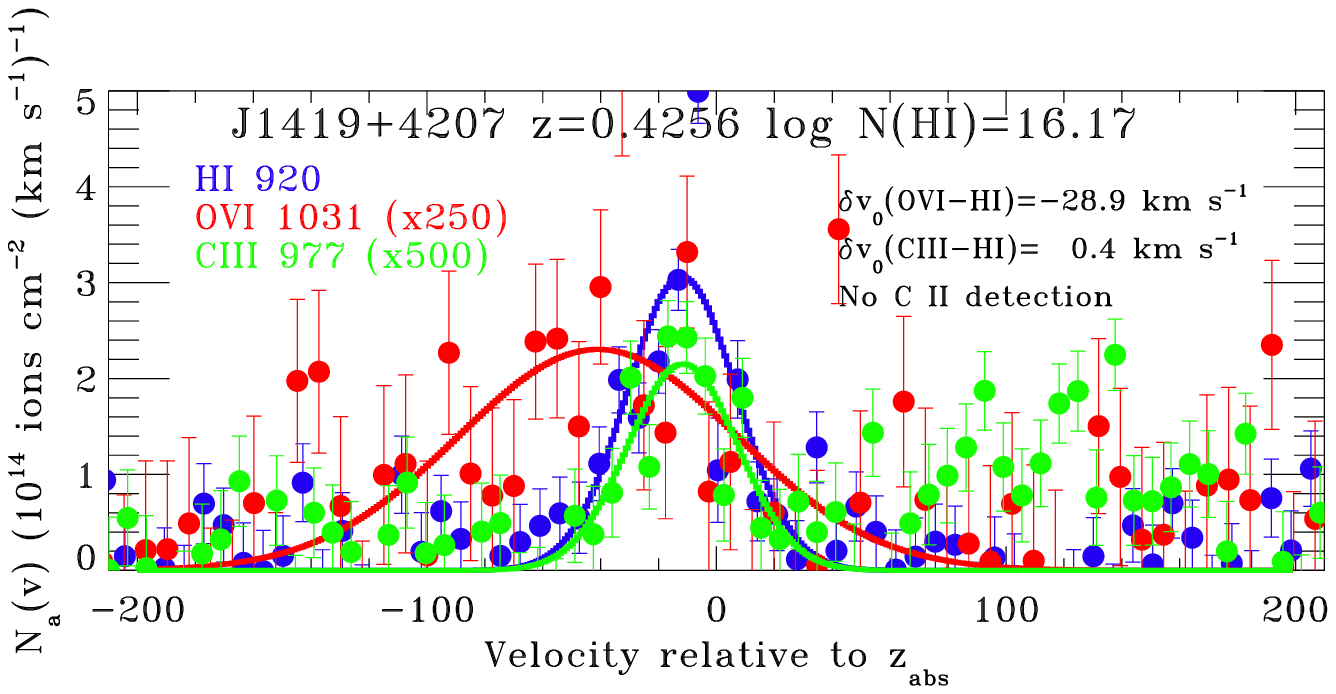}{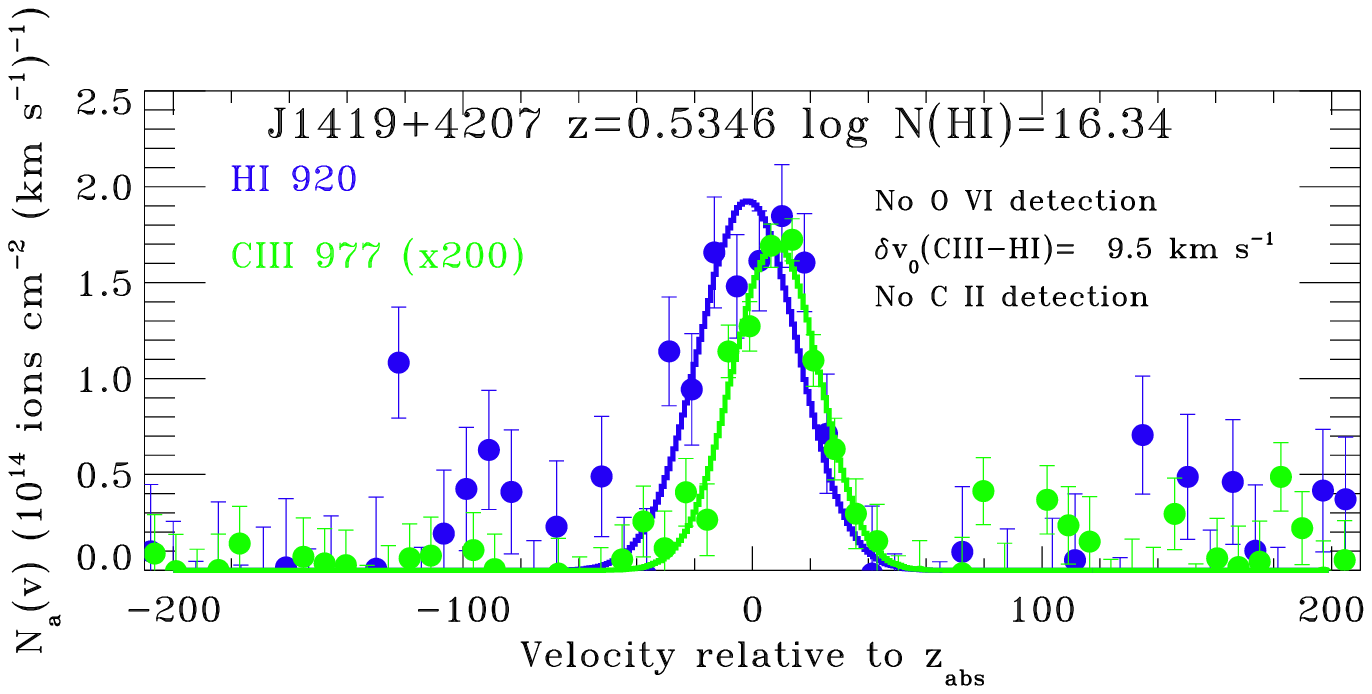}\end{figure}
\begin{figure}\epsscale{1.0}\plottwo{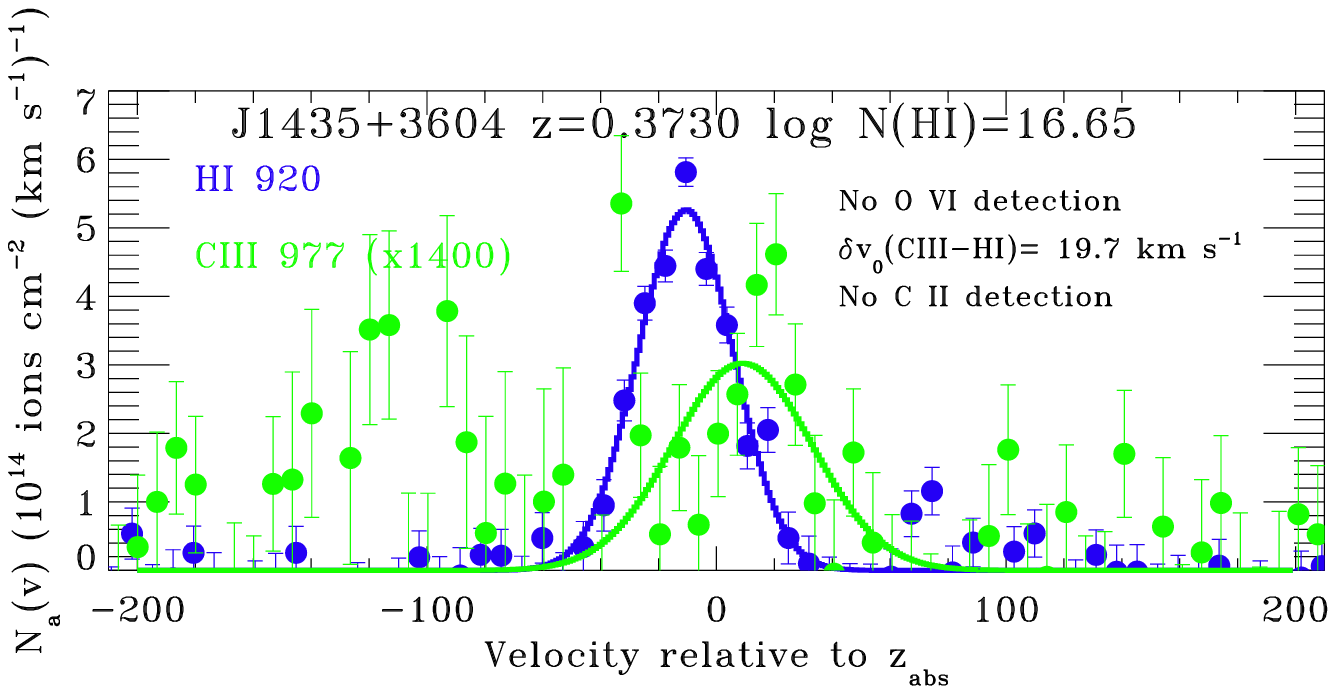}{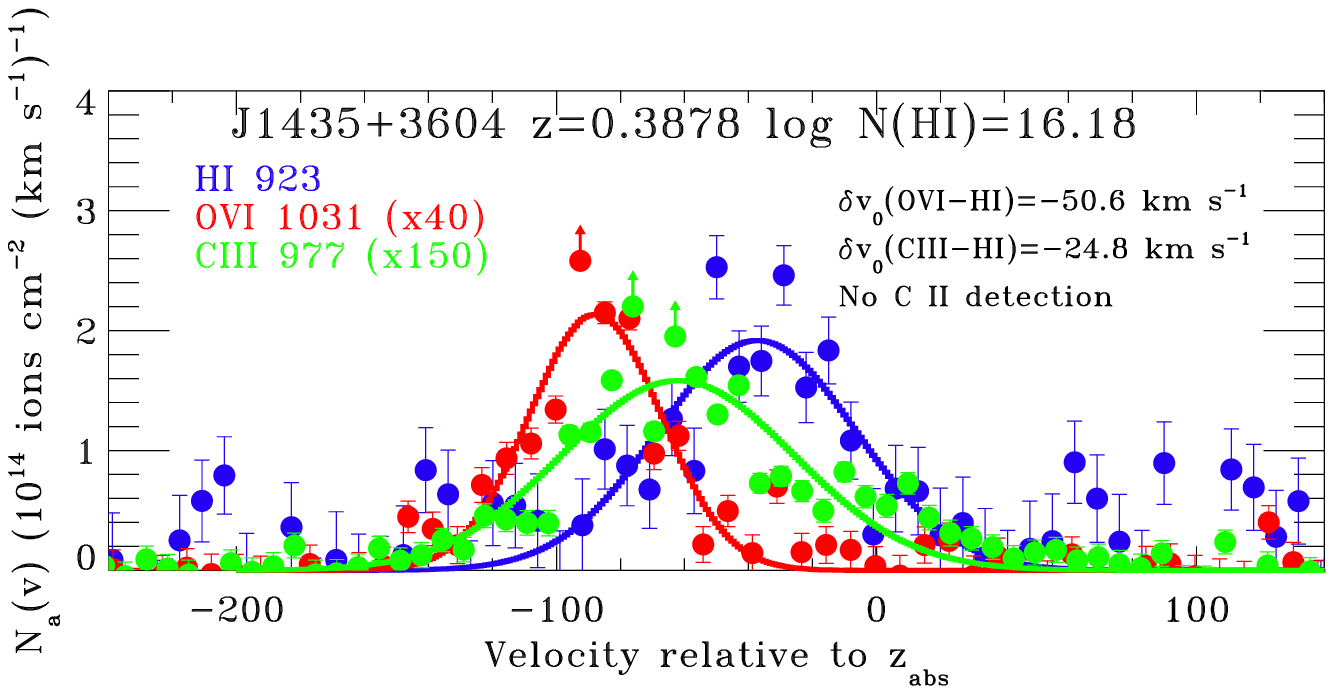}\end{figure}
\begin{figure}\epsscale{1.0}\plottwo{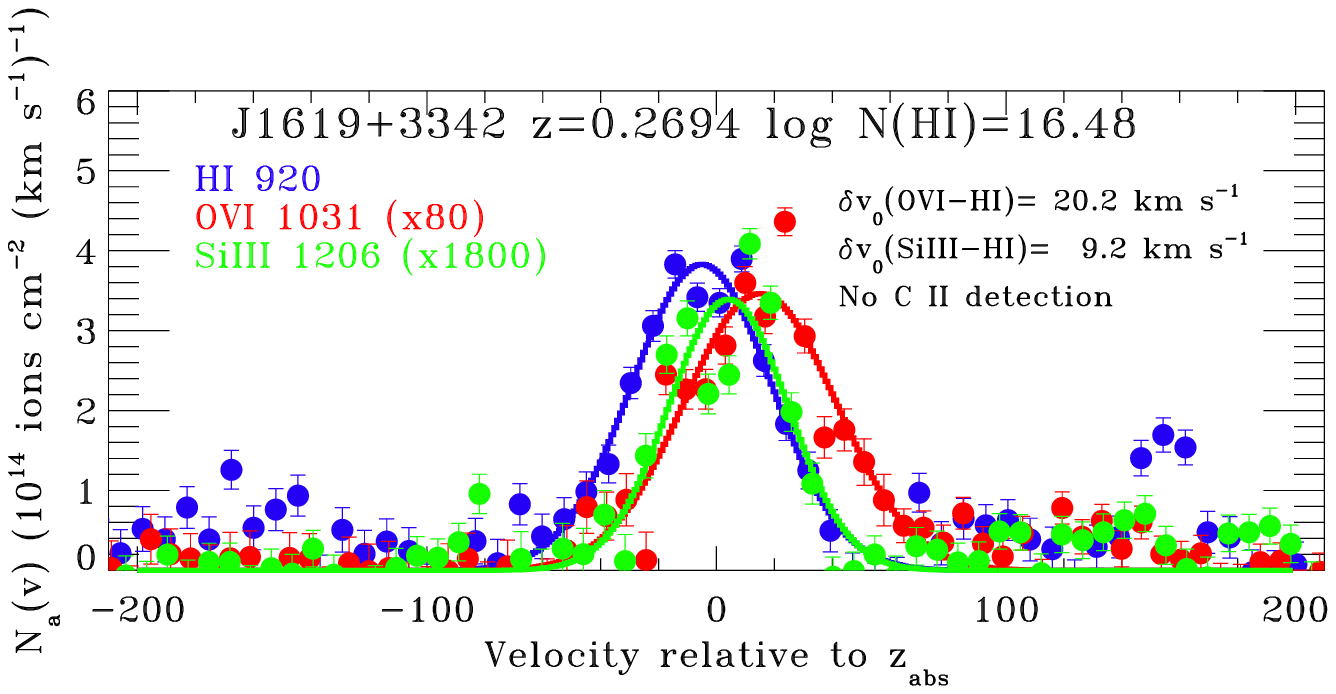}{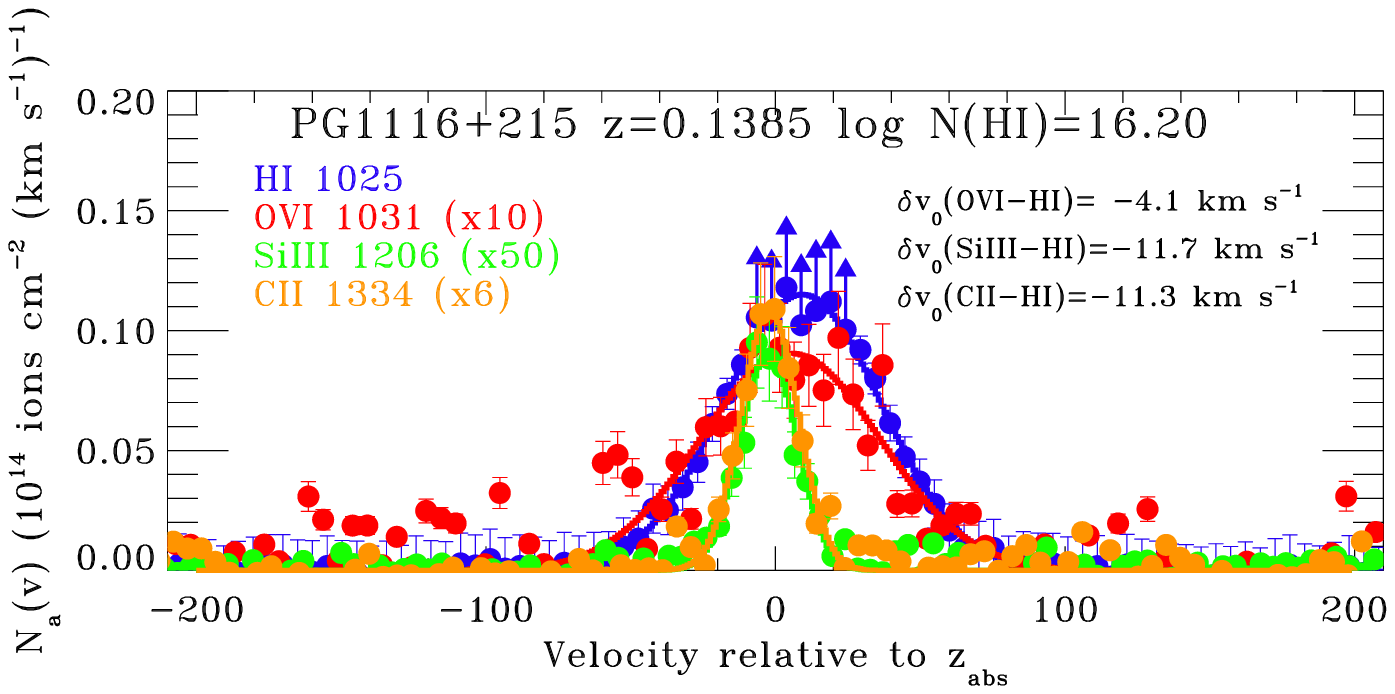}\end{figure}
\begin{figure}\epsscale{1.0}\plottwo{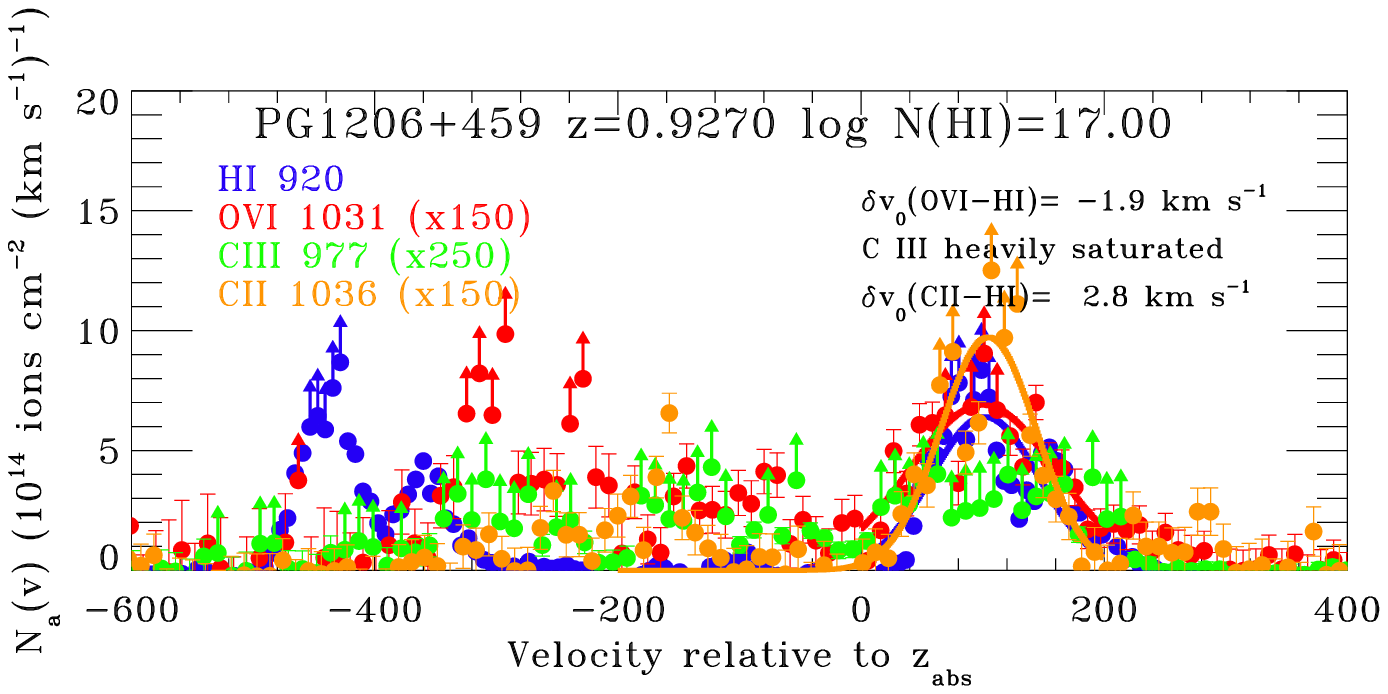}{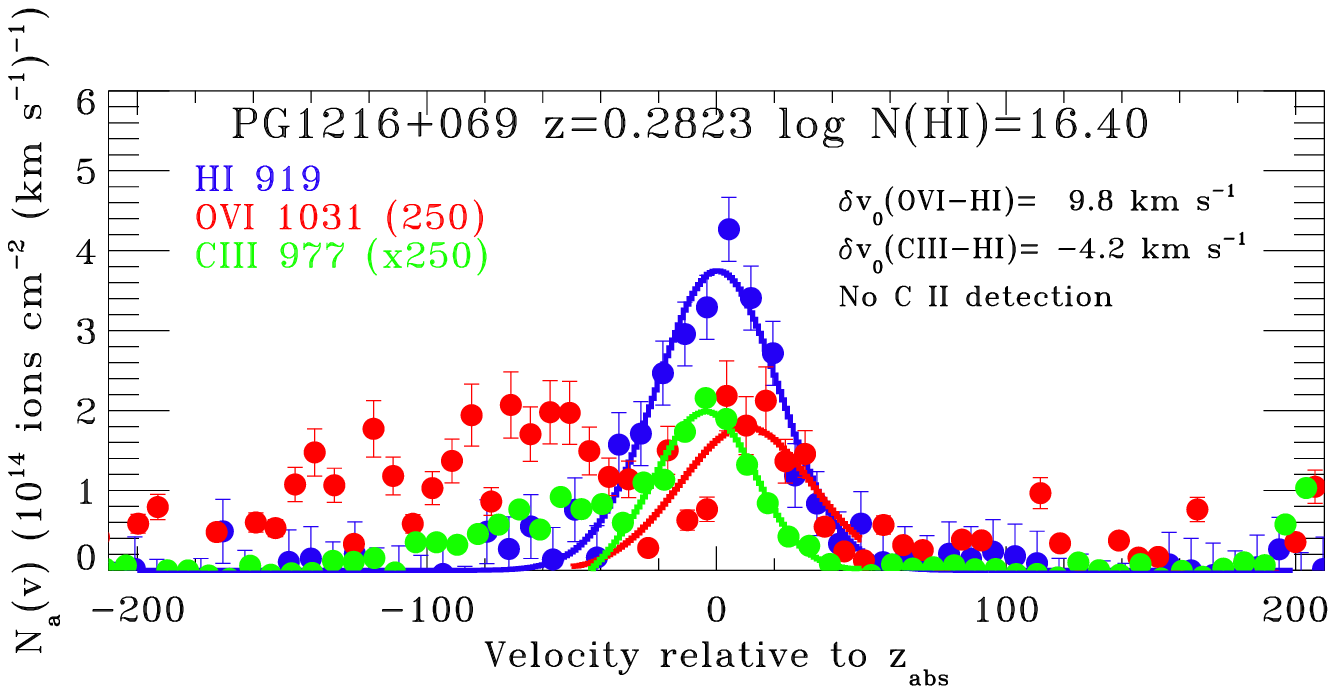}\end{figure}
\begin{figure}\epsscale{1.0}\plottwo{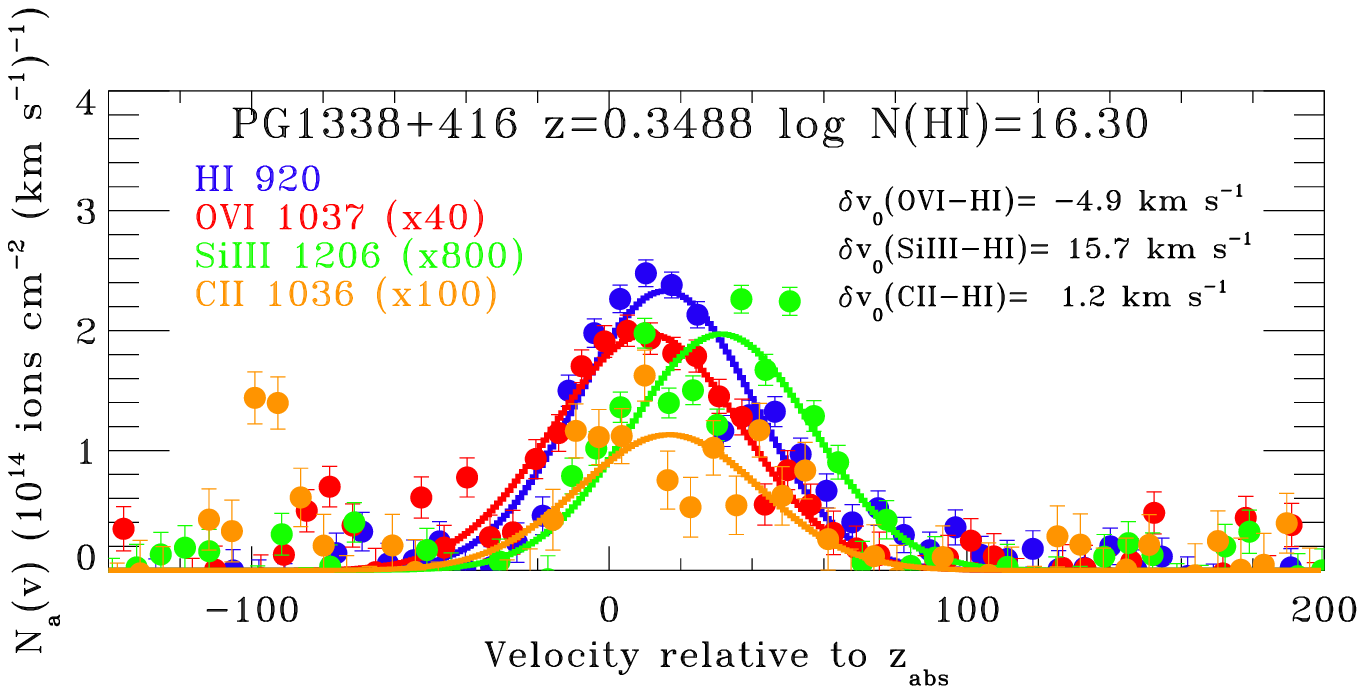}{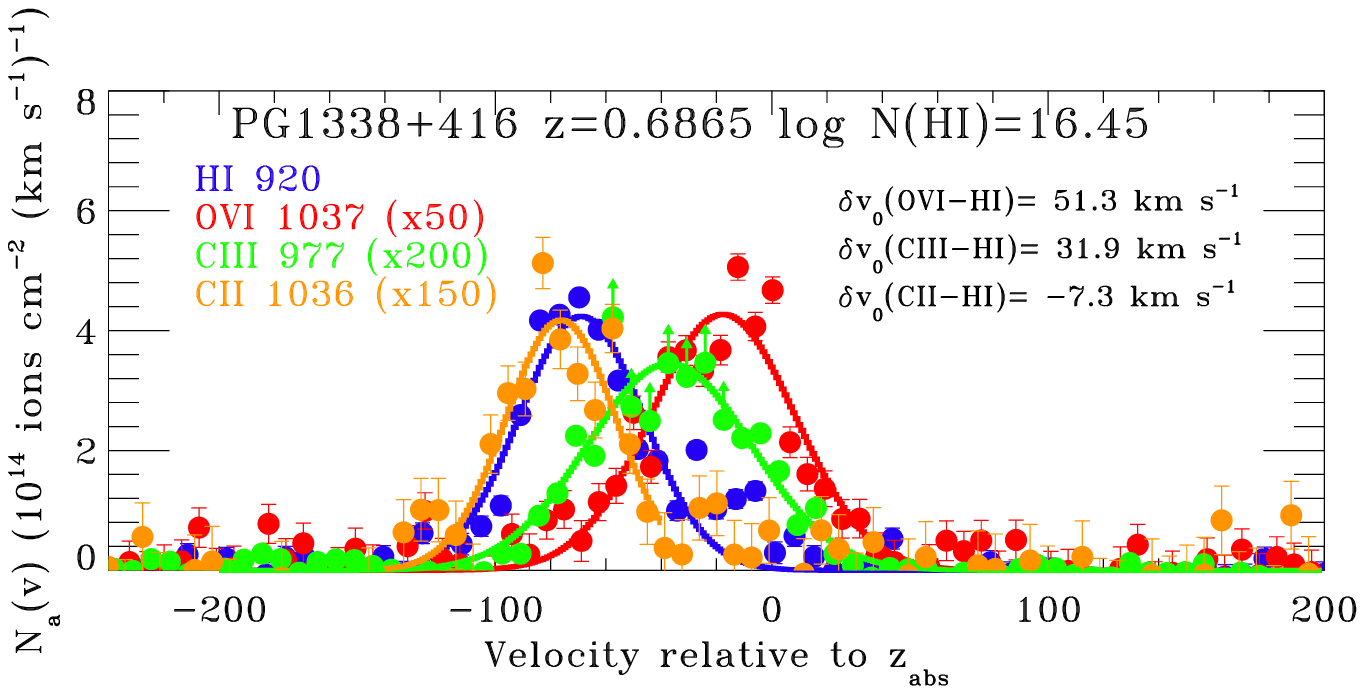}\end{figure}
\begin{figure}\epsscale{1.0}\plottwo{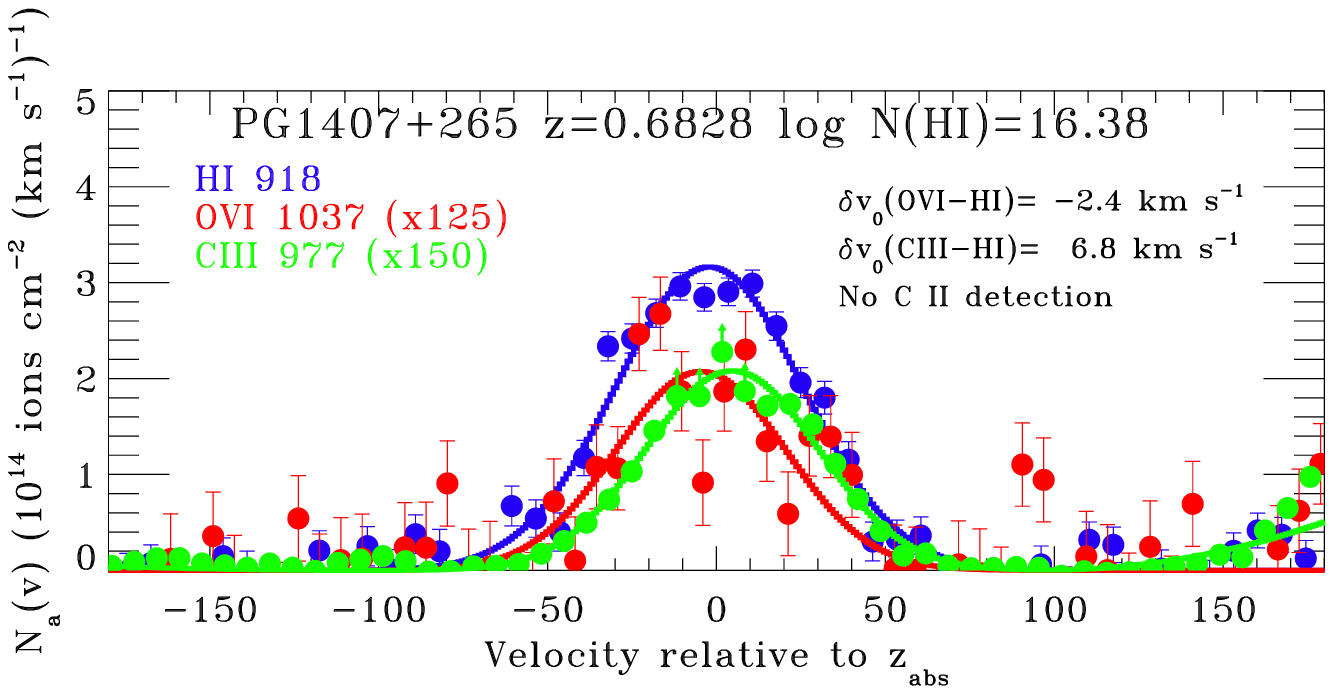}{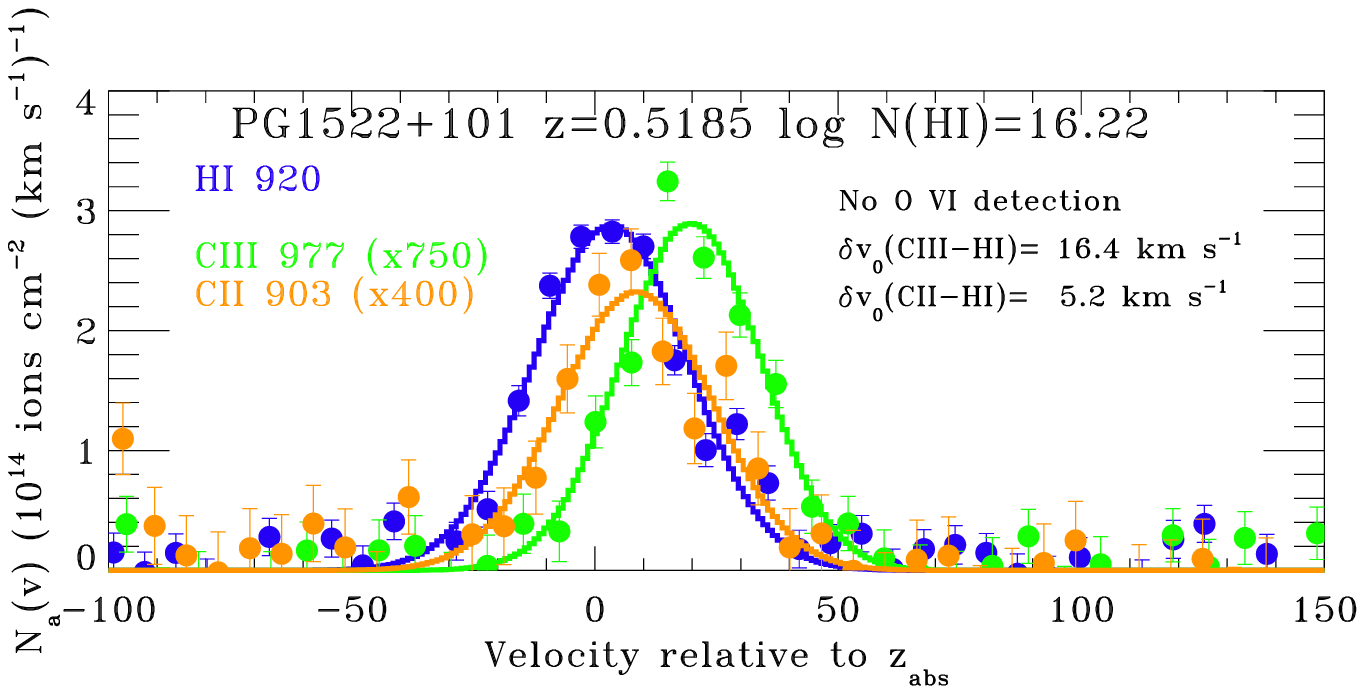}\end{figure}
\begin{figure}\epsscale{1.0}\plottwo{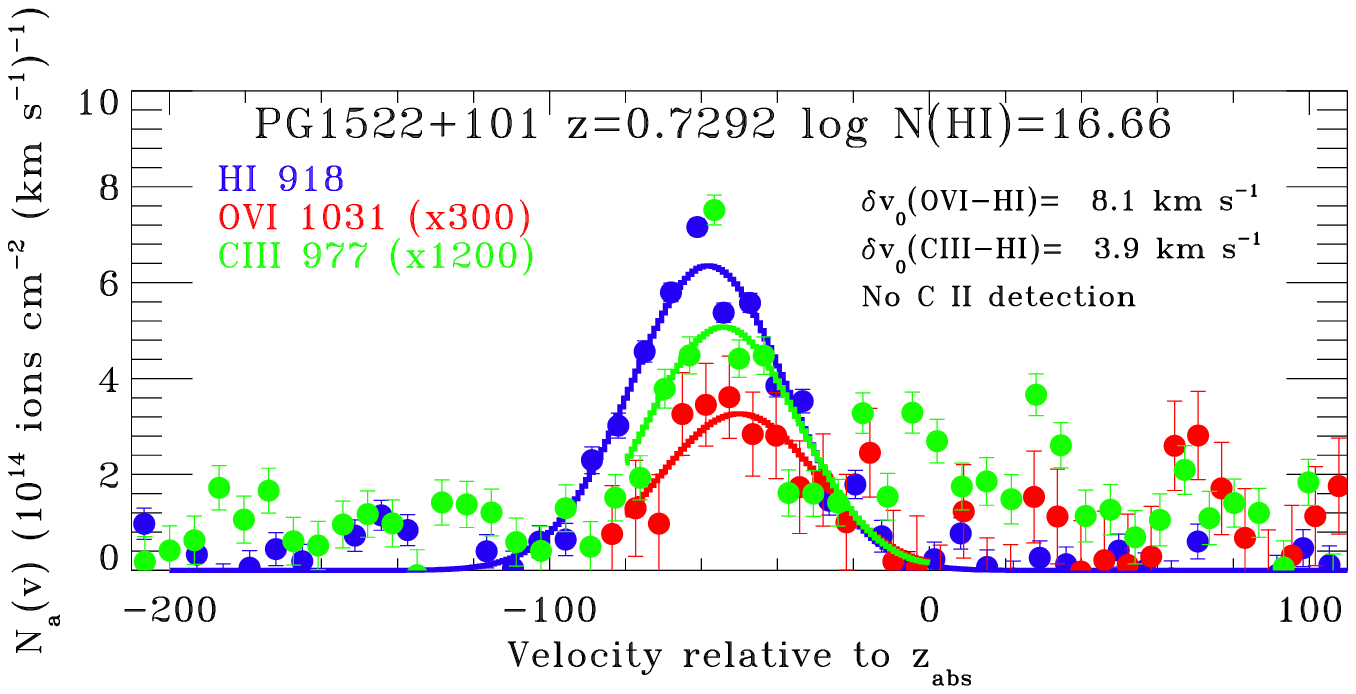}{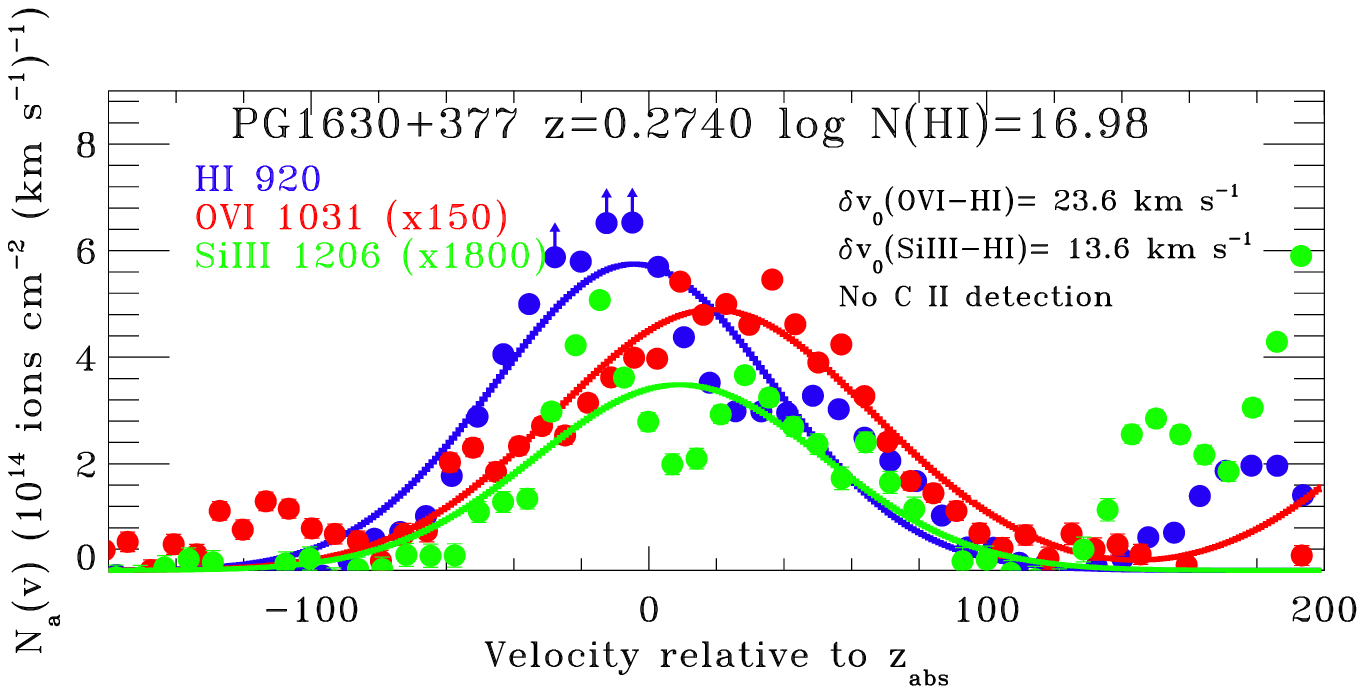}\end{figure}
\begin{figure}\epsscale{1.0}\plottwo{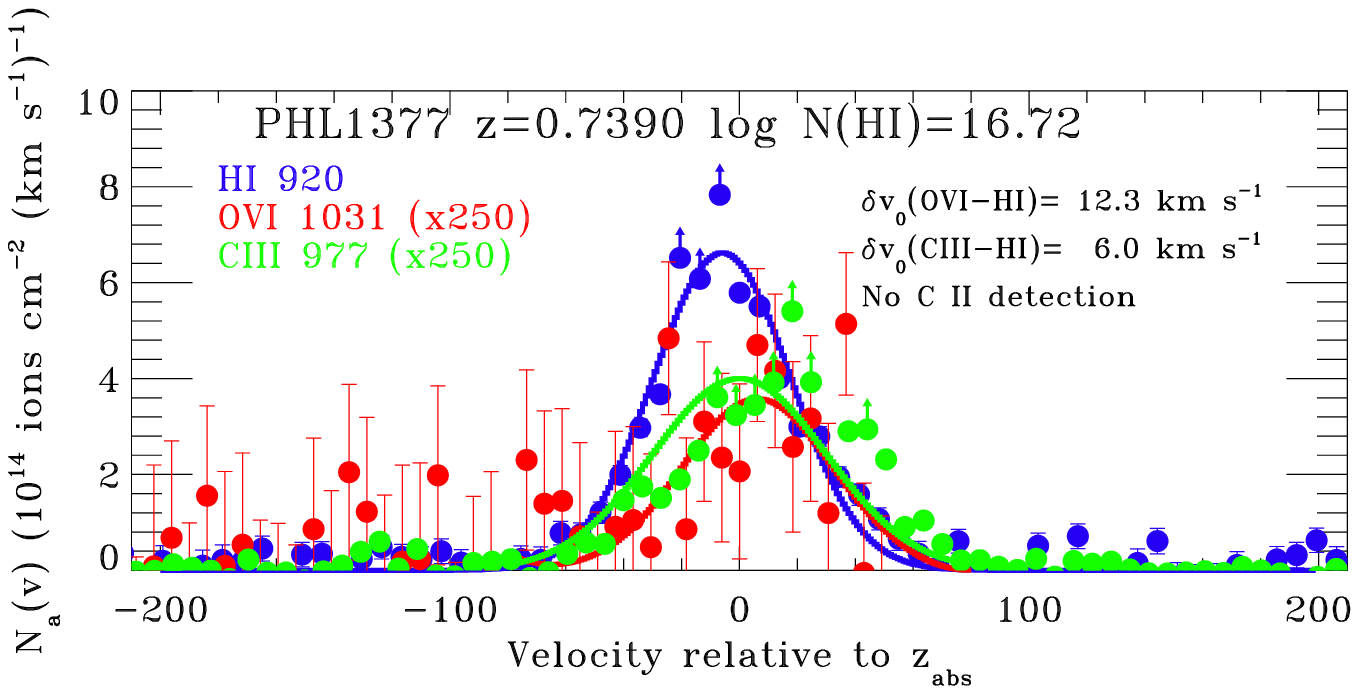}{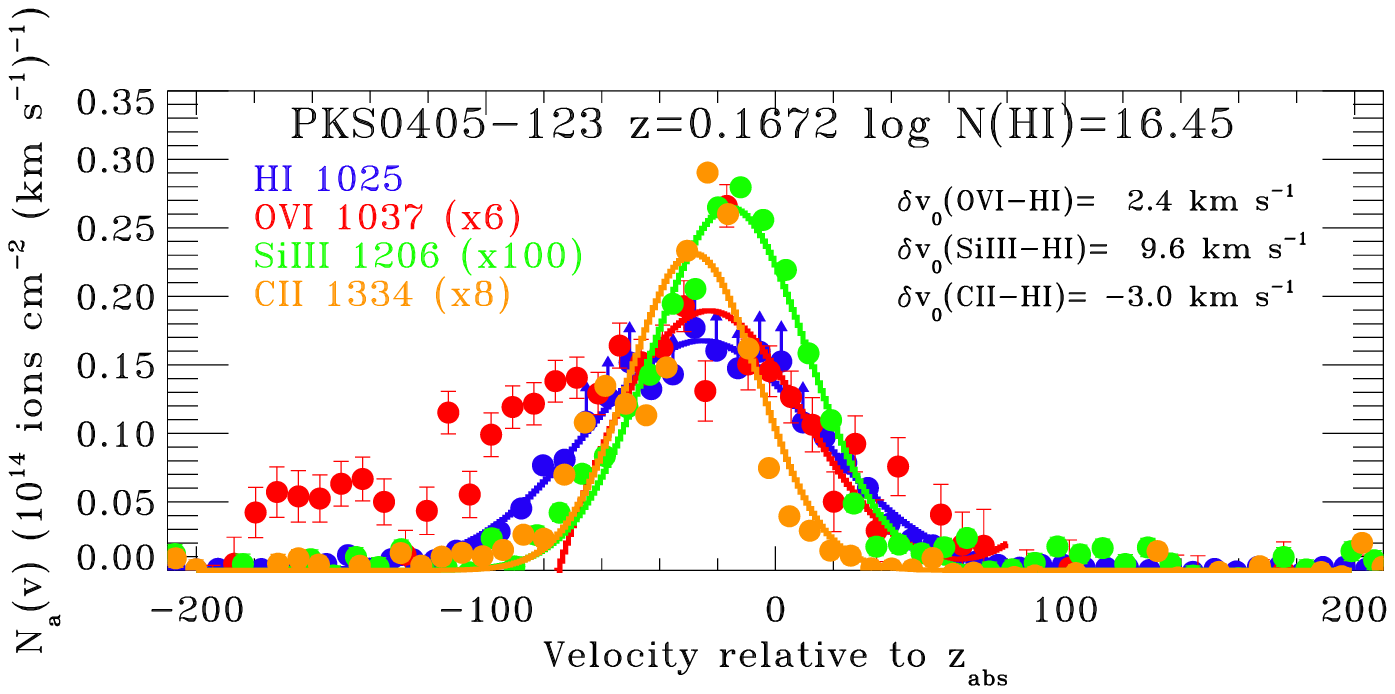}\end{figure}
\begin{figure}\epsscale{1.0}\plottwo{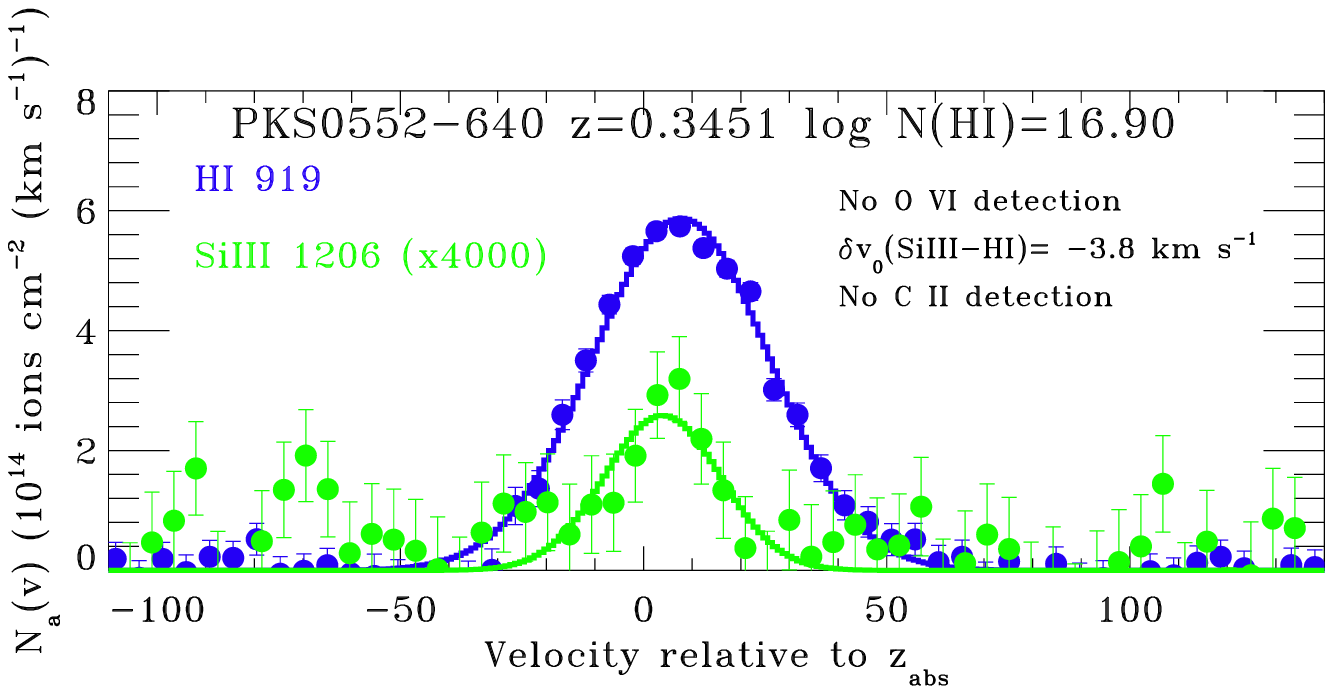}{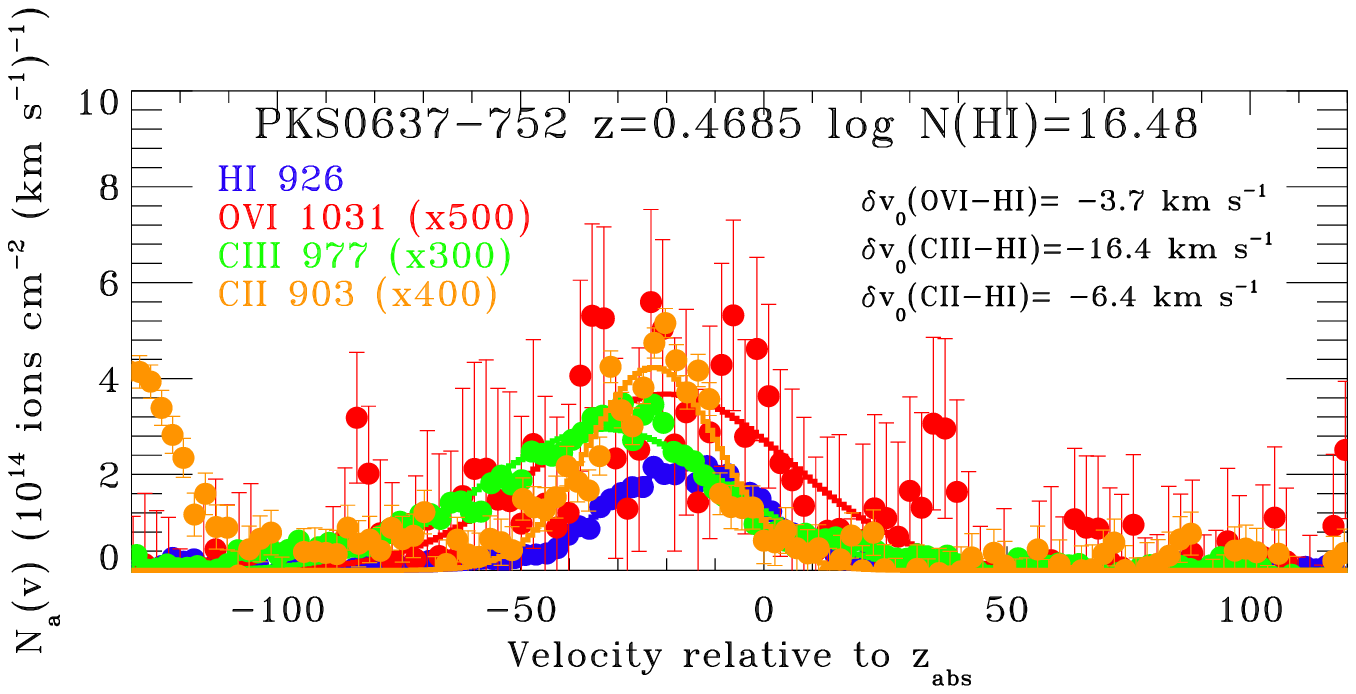}\end{figure}
\begin{figure}\epsscale{1.0}\plottwo{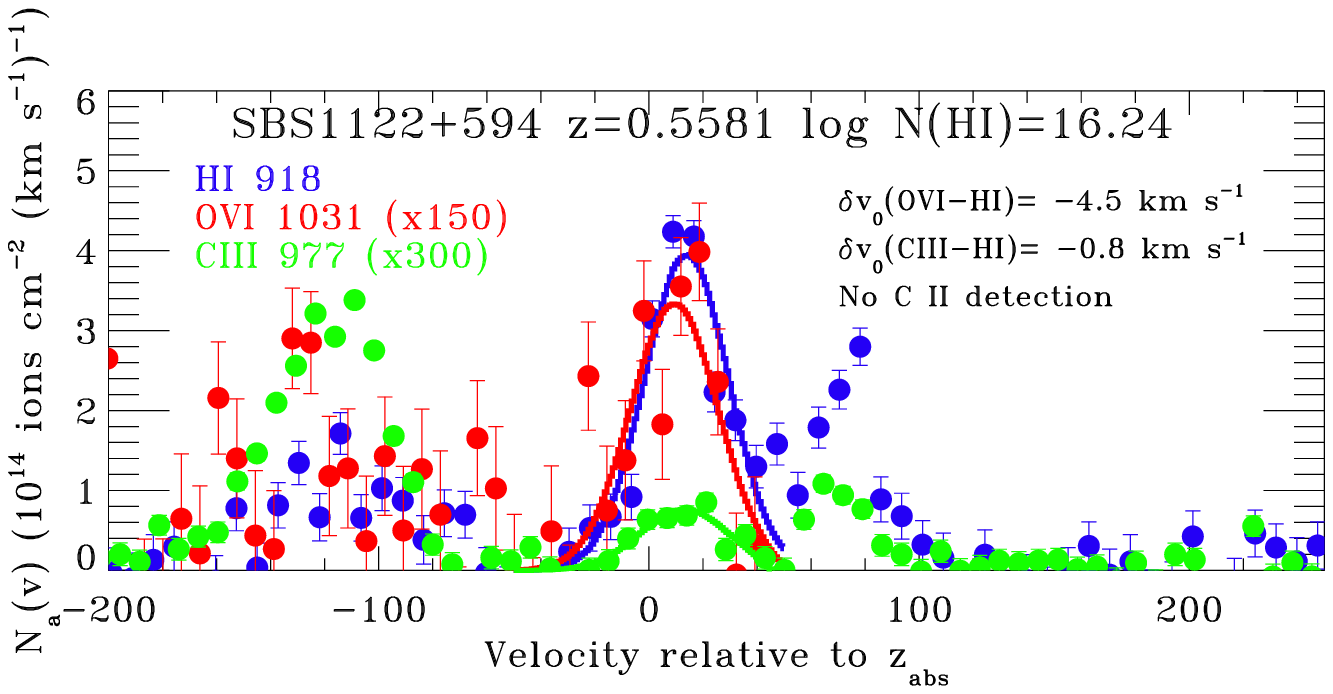}{nav.SBS1122+594.z=0.5581.eps}\end{figure}

\end{appendix}


\begin{deluxetable*}{lc ccccccc}
\tabletypesize{\tiny}
\tablewidth{0pt}
\tablecaption{LLS Column Densities}
\tablehead{Target & Prog. & $z_{\rm LLS}$ & [Z/H] & log\,$N$(\hi) & log\,$N$(\mgw) & log\,$N$(\cw) & log\,$N$(\ct) & log\,$N$(\os)\\
 & & & & (\sqcm) & (\sqcm) & (\sqcm) & (\sqcm) & (\sqcm)}
\startdata
    PG1522+101 &              11741 & 0.7292 &          $<\!-$2.0 &   16.66$\pm$0.05 &         $<$11.71 &                         $<$13.79 &                   13.33$\pm$0.03 &   13.72$\pm$0.07\\
    J1435+3604 &              11598 & 0.3730 &     $-$1.9$\pm$0.1 &   16.65$\pm$0.07 &   11.61$\pm$0.09 &                    $<$14.13$^\#$ &                   13.10$\pm$0.07 &         $<$13.52\\
    PG1407+265 &              11741 & 0.6828 &     $-$1.8$\pm$0.3 &   16.38$\pm$0.02 &         $<$11.96 &                    $<$13.17$^\#$ &                   13.94$\pm$0.01 &   13.99$\pm$0.06\\
    PG1630+377 &              11741 & 0.2740 &     $-$1.7$\pm$0.1 &   16.98$\pm$0.05 &   12.19$\pm$0.03 &                         $<$13.66 &            13.29$\pm$0.02$^\ast$ &   14.34$\pm$0.05\\
    PG1216+069 &              12025 & 0.2823 &          $<\!-$1.6 &   16.40$\pm$0.05 &          \nodata &                         $<$13.41 &                   13.69$\pm$0.01 &   13.91$\pm$0.05\\
    J1619+3342 &              11598 & 0.2694 &     $-$1.6$\pm$0.1 &   16.48$\pm$0.05 &   11.83$\pm$0.09 &                         $<$13.60 &            12.97$\pm$0.03$^\ast$ &   13.83$\pm$0.08\\
   PKS0552-640 &              11692 & 0.3451 &          $<\!-$1.5 &   16.90$\pm$0.08 &         $<$12.48 &                    $<$12.43$^\#$ &            12.33$\pm$0.08$^\ast$ &         $<$13.29\\
       PHL1377 &              11741 & 0.7390 &     $-$1.5$\pm$0.1 &   16.72$\pm$0.03 &   11.92$\pm$0.08 &                   12.99$\pm$0.12 &                   14.12$\pm$0.05 &   13.94$\pm$0.11\\
    J1419+4207 &              11598 & 0.4256 &     $-$1.4$\pm$0.2 &   16.17$\pm$0.06 &          \nodata &                         $<$13.71 &                   13.26$\pm$0.07 &   13.86$\pm$0.08\\
    J1435+3604 &              11598 & 0.3878 &          $<\!-$1.4 &   16.18$\pm$0.05 &          \nodata &                         $<$13.67 &                   14.02$\pm$0.03 &   14.14$\pm$0.06\\
    J0943+0531 &              11598 & 0.3544 &          $<\!-$1.3 &   16.11$\pm$0.09 &          \nodata &                         $<$14.09 &                   12.98$\pm$0.37 &         $<$14.06\\
   SBS1122+594 &              11520 & 0.5581 &     $-$1.0$\pm$0.1 &   16.24$\pm$0.03 &          \nodata &                         $<$14.13 &                   13.90$\pm$0.02 &         $<$14.11\\
   HE0153-4520 &              11541 & 0.2261 &     $-$0.8$\pm$0.2 &   16.61$\pm$0.15 &          \nodata &                   14.04$\pm$0.03 &            13.45$\pm$0.01$^\ast$ &   14.17$\pm$0.05\\
    PG1338+416 &              11741 & 0.3488 &     $-$0.6$\pm$0.2 &   16.30$\pm$0.13 &   12.46$\pm$0.04 &                   13.84$\pm$0.05 &            13.17$\pm$0.03$^\ast$ &   14.51$\pm$0.05\\
    J1419+4207 &              11598 & 0.2889 &     $-$0.6$\pm$0.1 &   16.40$\pm$0.07 &   12.56$\pm$0.03 &                   14.07$\pm$0.09 &                   13.97$\pm$0.04 &   14.53$\pm$0.05\\
    PG1116+215 &              12038 & 0.1385 &     $-$0.5$\pm$0.2 &   16.20$\pm$0.05 &          \nodata &                   13.70$\pm$0.04 &            12.68$\pm$0.03$^\ast$ &   13.85$\pm$0.05\\
   PKS0637-752 &              11692 & 0.4685 &     $-$0.5$\pm$0.1 &   16.48$\pm$0.04 &   12.63$\pm$0.08 &                   13.63$\pm$0.03 &                   13.81$\pm$0.01 &   13.66$\pm$0.08\\
    PG1522+101 &              11741 & 0.5185 &     $-$0.4$\pm$0.1 &   16.22$\pm$0.02 &   12.28$\pm$0.04 &                   13.34$\pm$0.05 &                   13.15$\pm$0.04 &         $<$13.33\\
   HE0439-5254 &              11520 & 0.6153 &     $-$0.3$\pm$0.1 &   16.28$\pm$0.04 &          \nodata &                   13.85$\pm$0.07 &                   14.41$\pm$0.06 &   14.87$\pm$0.05\\
    J1419+4207 &              11598 & 0.5346 &     $-$0.2$\pm$0.2 &   16.34$\pm$0.17 &   12.68$\pm$0.04 &                         $<$14.14 &                   13.52$\pm$0.06 &         $<$13.95\\
    PG1338+416 &              11741 & 0.6865 &       +0.1$\pm$0.1 &   16.45$\pm$0.05 &   12.93$\pm$0.01 &                   14.16$\pm$0.04 &                   14.09$\pm$0.06 &   14.76$\pm$0.05\\
   PKS0405-123 &              11508 & 0.1672 &       +0.1$\pm$0.2 &   16.45$\pm$0.05 &          \nodata &                   14.22$\pm$0.01 &            13.26$\pm$0.01$^\ast$ &   14.59$\pm$0.05\\
    PG1206+459 &    11741$^\dagger$ & 0.9270 &       +0.3$\pm$0.1 &   17.00$\pm$0.10 &          \nodata &                   14.94$\pm$0.08 &                   14.74$\pm$0.09 &         $>$15.19\\
\vspace{-0.25cm}
\enddata
\tablecomments{The redshift $z_{\rm LLS}$, metallicity [Z/H], and \hi\ column $N$(\hi) of each LLS are taken from L13. All other entries are new measurements. The systems are presented in order of increasing [Z/H]. Prog. refers to the \hst\ Program ID of the data used.The \os\ columns are derived from $\lambda$1031 unless saturated or blended, in which case $\lambda$1037 is used. The symbol $^\#$ indicates that \siw\ ($\lambda$1020 or $\lambda$1260) was used as a proxy for \cw. The symbol $^\ast$ indicates that \sit\ $\lambda$1206 was used as a proxy for \ct\ $\lambda$977. The symbol $^\dagger$ indicates that COS data from \hst\ program 12466 were also used.}
\end{deluxetable*}

\begin{deluxetable*}{l cc rrrrr rrr}
\tabletypesize{\tiny}
\tablewidth{0pt}
\tablecaption{LLS Kinematic Measurements}
\tablehead{Target & $z_{\rm LLS}$ & [Z/H] & \multicolumn{5}{c}{\underline{~~~~~~~~~~~~~$\Delta v_{90}$ (km\,s$^{-1}$)~~~~~~~~~~~~~}} & \multicolumn{3}{c}{\underline{~~~~~~~~~~~~$\delta v_0$ (km\,s$^{-1}$)~~~~~~~~~~~~}}\\
 & & & \hi\ & \mgw\ & \cw\ & \ct\ & \os\ & \os--\hi\ & \ct--\hi & \cw--\hi}
\startdata
    PG1522+101 &              0.7292 &          $<\!-$2.0 &                62 &         \nodata &         \nodata &              71 &              55 &                    8 &                    3 &              \nodata \\
    J1435+3604 &              0.3730 &     $-$1.9$\pm$0.1 &           49 &              11 &         \nodata &              66 &         \nodata &              \nodata &                   19 &              \nodata \\
    PG1407+265 &              0.6828 &     $-$1.8$\pm$0.3 &                78 &         \nodata &         \nodata &          $<$ 80 &              63 &                 $-$2 &                    6 &              \nodata \\
    PG1630+377 &              0.2740 &     $-$1.7$\pm$0.1 &          $<$129 &              33 &         \nodata &      114$^\ast$ &             108 &                   23 &            13$^\ast$ &              \nodata \\
    PG1216+069 &              0.2823 &          $<\!-$1.6 &                83 &         \nodata &         \nodata &             107 &             128 &                    9 &                 $-$4 &              \nodata \\
    J1619+3342 &              0.2694 &     $-$1.6$\pm$0.1 &               76 &              55 &         \nodata &       57$^\ast$ &              40 &                   20 &             9$^\ast$ &              \nodata \\
   PKS0552-640 &              0.3451 &          $<\!-$1.5 &                58 &         \nodata &         \nodata &       45$^\ast$ &         \nodata &              \nodata &          $-$3$^\ast$ &              \nodata \\
       PHL1377 &              0.7390 &     $-$1.5$\pm$0.1 &           $<$89 &              22 &              28 &          $<$ 90 &              67 &                   12 &              \nodata &              \nodata \\
    J1419+4207 &              0.4256 &     $-$1.4$\pm$0.2 &                68 &         \nodata &         \nodata &              51 &              67 &                $-$28 &                    0 &              \nodata \\
    J1435+3604 &              0.3878 &          $<\!-$1.4 &              146 &         \nodata &         \nodata &          $<$145 &              76 &                $-$50 &                $-$24 &              \nodata \\
    J0943+0531 &              0.3544 &          $<\!-$1.3 &               164 &         \nodata &         \nodata &         \nodata &         \nodata &              \nodata &              \nodata &              \nodata \\
   SBS1122+594 &              0.5581 &     $-$1.0$\pm$0.1 &              215 &         \nodata &         \nodata &             238 &         \nodata &                 $-$4 &                 $-$0 &              \nodata \\
   HE0153-4520 &              0.2261 &     $-$0.8$\pm$0.2 &           $<$85 &         \nodata &              73 &   $<$ 74$^\ast$ &             106 &                   12 &          $-$1$^\ast$ &                    3 \\
    PG1338+416 &              0.3488 &     $-$0.6$\pm$0.2 &             93 &              54 &              70 &       60$^\ast$ &              89 &                 $-$4 &            15$^\ast$ &                    1 \\
    J1419+4207 &              0.2889 &     $-$0.6$\pm$0.1 &              143 &              50 &             102 &          $<$ 92 &             101 &                   25 &                    8 &                $-$12 \\
    PG1116+215 &              0.1385 &     $-$0.5$\pm$0.2 &           $<$92 &         \nodata &              53 &       43$^\ast$ &             106 &                 $-$4 &         $-$11$^\ast$ &                $-$11 \\
   PKS0637-752 &              0.4685 &     $-$0.5$\pm$0.1 &           61 &              87 &              87 &             102 &              87 &                 $-$3 &                $-$16 &                 $-$6 \\
    PG1522+101 &              0.5185 &     $-$0.4$\pm$0.1 &            57 &              20 &              52 &              51 &         \nodata &              \nodata &                   16 &                    5 \\
   HE0439-5254 &              0.6153 &     $-$0.3$\pm$0.1 &             163 &         \nodata &              32 &          $<$195 &             177 &                    3 &                 $-$2 &                 $-$8 \\
    J1419+4207 &              0.5346 &     $-$0.2$\pm$0.2 &          54 &              31 &         \nodata &              58 &         \nodata &              \nodata &                    9 &              \nodata \\
    PG1338+416 &              0.6865 &       +0.1$\pm$0.1 &              92 &          $<$ 45 &              75 &          $<$ 86 &              94 &                   51 &                   31 &                 $-$7 \\
   PKS0405-123 &              0.1672 &       +0.1$\pm$0.2 &           $<$127 &         \nodata &              77 &       93$^\ast$ &             177 &                    2 &             9$^\ast$ &                 $-$2 \\
    PG1206+459 &              0.9270 &       +0.3$\pm$0.1 &            $<$345 &         \nodata &          $<$413 &          $<$533 &          $<$500 &                 $-$1 &              \nodata &                    2 \\
\vspace{-0.25cm}
\enddata
\tablecomments{The redshift $z_{\rm LLS}$ and metallicity [Z/H] of each LLS are taken from L13. All other entries are new measurements. The systems are presented in order of increasing [Z/H]. The errors on the velocity widths $\Delta v_{90}$ and velocity centroid offsets $\delta v_0$ are each $\approx$10\kms. No entry is given for $\Delta v_{90}$ or $\delta v_0$ if the relevant line is undetected. For saturated lines, upper limits on $\Delta v_{90}$ are given. The symbol $^\#$ indicates that \siw\ ($\lambda$1020 or $\lambda$1260) was used as a proxy for \cw. The symbol $^\ast$ indicates that \sit\ $\lambda$1206 was used as a proxy for \ct\ $\lambda$977. }
\end{deluxetable*}

\end{document}